\documentstyle[12pt,epsfig]{article}

\begin{document} 
\pagestyle{empty}

\begin{center} 
{\Large The Minimal Supersymmetric Universal Seesaw Mechanism (MSUSM). }
\end{center}

\begin{center}
M. C. Rodriguez \\
{\it Grupo de F\'\i sica Te\'orica e Matem\'atica F\'\i sica \\
Departamento de F\'\i sica \\
Universidade Federal Rural do Rio de Janeiro - UFRRJ \\
BR 465 Km 7, 23890-000 \\
Serop\'edica, RJ, Brazil, \\
email: marcoscrodriguez@ufrrj.br \\} 
\end{center} 

\begin{abstract}
We build a supersymmetric model with 
$SU(2)_{L}\otimes SU(2)_{R}\otimes U(1)_{(B-L)}$ electroweak gauge symmetry, where $SU(2)_{L}$ is the left-handed currents while $SU(2)_{R}$ is the right-handed currents and $B$ and $L$ are the usual baryonic and leptonic numbers. We can generate an universal seesaw mechanism to get masses for all the usual fermions in this model, it means quarks and leptons, and also explain the mixing experimental data. We will also to study the masses of the Gauge Bosons and also the masses of all usual scalars of this model.
\end{abstract}
PACS number(s): 12.60. Jv.


\section{Introduction}
\label{sec:intro}

Today one of the most popular extension of the Standard Model is the Minimal 
Supersymmetric Standard Model 
(MSSM) (see \cite{Rodriguez:2019mwf} and references therein). There are also the 
Minimal Supersymmetric Standard Model with three right-handed neutrinos (MSSM3RHN) and some Supersymmetric $(B-L)$ models, where the type-I seesaw mechanism is implemented for the generating 
masses for all the neutrinos; there are good candidates 
for Dark Matter and due a Majorana phase at sneutrinos masses it is 
possible to induce Leptogenesis in this model (see \cite{Rodriguez:2016esw,Rodriguez:2020} 
and references therein).

However, the left-right symmetric model 
have as basic assumption the following 
electroweak gauge symmetry
\begin{equation}
SU(2)_{L}\otimes SU(2)_{R}\otimes 
U(1)_{(B-L)}\otimes {\cal P}\otimes 
{\cal C},
\end{equation}
where both left-handed currents and right-handed currents must exist, and the former ones must be suppressed. There are also version where the parity ${\cal P}$ or the 
charge conjugation ${\cal C}$ are not introduced in the model\footnote{See also their discussion and references for more details about this kind of models.} \cite{Diaz:2020qti}.

There are another interesting left-right symmetric model, it is known as the 
``Left-Right Universal Seesaw 
Mechanism (LRUSM), presented at \cite{Davidson:1987mh,
Davidson:1987tr}. It is an interesting extension of the Standard Model because on this model the charged fermions, as well the neutrinos, get their masses tnrough a seesaw mechanism, and this mechanism is called 
as universal seesaw mechanism. In the LRUSM, 
the scalar sector is very simple when compared with the usual left-right model, where you need at least to introduce bidoublet scalar field to generate mass for all the fermions in the model.

We will present this model with only one 
fermion familly, as done at \cite{Davidson:1987mh} because our main interest is in study the masses of all Gauge bosons masses and the masses in the scalar sector to show that they are in concordance with all the experimental data.

This paper is organized as follows. In 
Sec. \ref{sec:particles} we present our model 
and then we built our lagrangian using superfield 
formalism. We present the masses for all the 
leptons, gauge bosons and the usual scalar mass spectrum. Finally, the last section is
devoted to our conclusions.

\section{The Model}
\label{sec:particles}

Our gauge symmetry is defined  in the same way as in the LRUSM, it means the following 
gauge group
\begin{equation}
SU(3)_{C}\otimes SU(2)_{L}\otimes SU(2)_{R}\otimes 
U(1)_{(B-L)}.
\end{equation} 
there are several supersymmetric left-right models in the literature 
\cite{susylr,doublet}.

The gauge bosons and their superpartners, known as gauginos, 
they are put in vector superfields as shown at Tab.(\ref{table1}) and we have 
to impose the following constraints \cite{Davidson:1987mh,
Davidson:1987tr}:
\begin{equation}
g_{L}\neq g_{R},
\end{equation}
and Parity Symmetry is present at any energy scale, 
because we assume that left-right symmetry is explicitly 
broken.

\begin{table}[h]
\begin{center}
\begin{tabular}{|c|c|c|c|c|c|}
\hline
${\rm{Group}}$ & ${\rm Superfield}$ & ${\rm{Bosons}}$ & ${\rm{Gaugino}}$ & ${\rm{Auxiliar \,\ Field}}$ & ${\rm constant}$ \\
\hline
$SU(3)_{C}$ & $\hat{G}^{a}$ & $g^{a}_{m}$ & $\tilde{g}^{a}$ & $D^{a}_{g}$ & $g_{s}$  \\
\hline
$SU(2)_{L}$ & $\hat{V}^{i}_{L}$ & $(V^{i}_{L})_{m}$ & $\tilde{V}^{i}_{L}$ & $D^{i}_{L}$ & $g_{L}$ \\
\hline
$SU(2)_{R}$ & $\hat{V}^{i}_{R}$ & $(V^{i}_{R})_{m}$ & $\tilde{V}^{i}_{R}$ & $D^{i}_{R}$ & $g_{R}$ \\
\hline
$U(1)_{(B-L)}$ & $\hat{V}^{BL}$ & $V^{BL}_{m}$ & $\tilde{V}^{BL}$ & $D^{BL}$ & $g_{BL}$ \\
\hline
\end{tabular}
\end{center}
\caption{Information on fields contents of each vector superfield of this model. The Latin index $m$ identify Lorentz index, 
while $a=1,2, \ldots ,8$ and $i=1,2,3$.}
\label{table1}
\end{table}

The eletric charge operator is defined as usual
\begin{equation}
\frac{Q}{e}=T_{3L}+T_{3R}+ 
\frac{(B-L)}{2},
\end{equation}
where $T_{3L}$ and $T_{3R}$ are, respectively, the third 
component of isospin of the 
gauge groups $SU(2)_{L}$ and 
$SU(2)_{R}$ and $(B-L)$ is the diference between baryon (B) and lepton (L) number.

We will focus on one family 
case. We introduce the 
followings fermions at 
LRUSM at doublet representation \cite{Davidson:1987mh,
Davidson:1987tr}:
\begin{eqnarray}
L_{L} &=&\left( 
\begin{array}{c}
\nu_{e} \\ 
e
\end{array}
\right)_{L}\sim \left( {\bf 1},{\bf 2},{\bf 1},-1 \right) , \,\ 
L_{R} =\left( 
\begin{array}{c}
e^{+} \\ 
- \nu^{c}_{e}
\end{array}
\right)_{R}\sim \left( {\bf 1},{\bf 1},{\bf 2}^{*},+1 \right) ,  \nonumber \\
Q_{L} &=&\left( 
\begin{array}{c}
u \\ 
d
\end{array}
\right)_{L}\sim \left( {\bf 3},{\bf 2},{\bf 1},+ \frac{1}{3} \right) , \,\ 
Q_{R}= \left( 
\begin{array}{c}
d^{c} \\ 
-u^{c}
\end{array}
\right)_{R}\sim \left( {\bf 3^{*}},{\bf 1},{\bf 2}^{*},- \frac{1}{3}\right),  \nonumber \\
\label{fermionsd}
\end{eqnarray}
and at singlet representation \cite{Davidson:1987mh,
Davidson:1987tr}:
\begin{eqnarray}
E^{H}_{L}&\sim& 
\left( {\bf 1},{\bf 1},
{\bf 1},-2 \right) , \,\
E^{H}_{R}\sim 
\left( {\bf 1},{\bf 1},
{\bf 1},+2 \right) , 
\nonumber \\ 
N^{H}_{L}&\sim& 
\left( {\bf 1},{\bf 1},
{\bf 1},0 \right) , 
\,\
N^{H}_{R}\sim 
\left( {\bf 1},{\bf 1},
{\bf 1},0 \right) , 
\nonumber \\
U^{H}_{L}&\sim& 
\left( {\bf 3},{\bf 1},
{\bf 1},+ \frac{4}{3} \right) , \,\ 
U^{H}_{R}\sim 
\left( {\bf 3^{*}},{\bf 1},{\bf 1},+ \frac{4}{3} \right) , \nonumber \\
D^{H}_{L}&\sim& \left( {\bf 3},{\bf 1},{\bf 1},- \frac{2}{3} \right) , \,\
 \,\ 
D^{H}_{R}\sim \left( {\bf 3^{*}},{\bf 1},{\bf 1},+ 
\frac{2}{3} \right). \nonumber \\
\label{fermions}
\end{eqnarray}

In supersymmetric model we 
introduce the particles defined at Eq.(\ref{fermions}) in the chiral superfields defined at 
Tabs.(\ref{lep},\ref{quark}).

\begin{table}[h]
\begin{center}
\begin{tabular}{|c|c|c|c|}
\hline
$\mbox{ Chiral Superfield} $ &  $\mbox{ Sleptons} $ & $\mbox{ Leptons} $ & ${\rm{Auxiliar \,\ Field}}$ \\ \hline
$\hat{L}_{L}$ & $\tilde{L}_{L}$ & $L_{L}$ &  $F_{L_{L}}$ \\ \hline
$\hat{L}_{R}$ & $\tilde{L}_{R}$ & $L_{R}$ &  $F_{L_{R}}$ \\ \hline
$\hat{E}^{H}_{L}$ & $\tilde{E}^{H}_{L}$ & $E^{H}_{L}$ &  $F_{E^{H}_{L}}$ \\ \hline
$\hat{N}^{H}_{L}$ & $\tilde{N}^{H}_{L}$ & $N^{H}_{L}$ &  $F_{N^{H}_{L}}$  \\ \hline
$\hat{E}^{H}_{R}$ & $\tilde{E}^{H}_{R}$ & $E^{H}_{R}$ &  $F_{E^{H}_{R}}$ \\ \hline
$\hat{N}^{H}_{R}$ & $\tilde{N}^{H}_{R}$ & $N^{H}_{R}$ &  $F_{N^{H}_{R}}$  \\ \hline
\end{tabular}
\end{center}
\caption{Particle content in each chiral superfield defined for the leptons defined in Eq.(\ref{fermions}).}
\label{lep}
\end{table}

\begin{table}[h]
\begin{center}
\begin{tabular}{|c|c|c|c|}
\hline
$\mbox{ Chiral Superfield} $ &  $\mbox{ Squarks} $ & $\mbox{ Quarks} $ & ${\rm{Auxiliar \,\ Field}}$ \\ \hline
$\hat{Q}_{L}$ & $\tilde{Q}_{L}$ & $Q_{L}$ &  $F_{Q_{L}}$ \\ \hline
$\hat{Q}_{R}$ & $\tilde{Q}_{R}$ & $Q_{R}$ &  $F_{Q_{R}}$ \\ \hline
$\hat{U}^{H}_{L}$ & $\tilde{U}^{H}_{L}$ & $U^{H}_{L}$ &  $F_{U^{H}_{L}}$ \\ \hline
$\hat{D}^{H}_{L}$ & $\tilde{D}^{H}_{L}$ & $N^{D}_{L}$ &  $F_{D^{H}_{L}}$  \\ \hline
$\hat{U}^{H}_{R}$ & $\tilde{U}^{H}_{R}$ & $U^{H}_{R}$ &  $F_{U^{H}_{R}}$ \\ \hline
$\hat{D}^{H}_{R}$ & $\tilde{D}^{H}_{R}$ & $D^{H}_{R}$ &  $F_{D^{H}_{R}}$  \\ \hline
\end{tabular}
\end{center}
\caption{Particle content in each chiral superfield defined for the quarks defined in Eq.(\ref{fermions}).}
\label{quark}
\end{table}

The associated Higgs system at Universal Seesaw mechanism is defined as \cite{Davidson:1987mh}
\begin{eqnarray}  
\phi_{L} &=&\left( 
\begin{array}{c}
\phi^{0}_{L} \\ 
\phi^{-}_{L}
\end{array}
\right)_{L}\sim \left( {\bf 1},{\bf 2},{\bf 1},-1 \right) , \,\ 
\phi_{R} =\left( 
\begin{array}{c}
\phi^{0}_{R} \\ 
\phi^{-}_{R}
\end{array}
\right)_{R}\sim \left( {\bf 1},{\bf 1},{\bf 2},-1 \right) .
\label{usualscalars}
\end{eqnarray}

As in the Minimal Supersymmetric Standard Model, in order to avoid anomalies and give mass to all the fermions, we need to 
introduce the followings new 
scalars:
\begin{eqnarray}  
\phi^{\prime}_{L} &=&\left( 
\begin{array}{c}
\phi^{\prime +}_{L} \\ 
\phi^{\prime 0}_{L}
\end{array}
\right)_{L}\sim \left( {\bf 1},{\bf 2^{*}},{\bf 1},+1 \right) , \,\ 
\phi^{\prime}_{R} =\left( 
\begin{array}{c}
\phi^{\prime +}_{R} \\ 
\phi^{\prime 0}_{R}
\end{array}
\right)_{R}\sim \left( {\bf 1},{\bf 1},{\bf 2^{*}},+1 \right) ,
\label{newscalars}
\end{eqnarray}
in order to get the $\chi$-Term, defined at Eq.(11) \cite{Davidson:1987mh}, we have to add an extra singlet defined as 
\begin{equation}
S \sim ({\bf 1},{\bf 1},{\bf 1},0).
\label{newsingletscalar}
\end{equation} 
This singlet was also considered at \cite{Davidson:1987tr}, as we will present at Sec.(\ref{sec:sp}).

\begin{table}[h]
\begin{center}
\begin{tabular}{|c|c|c|c|}
\hline
$\mbox{ Chiral Superfield} $ &  $\mbox{ Scalars} $ & $\mbox{ Higgsinos} $ & ${\rm{Auxiliar \,\ Field}}$ \\ \hline
$\hat{\phi}_{L}$ & $\phi_{L}$ & $\tilde{\phi}_{L}$ &  $F_{\phi_{L}}$ \\ \hline
$\hat{\phi}_{R}$ & $\phi_{R}$ & $\tilde{\phi}_{R}$ & $F_{\phi_{R}}$ \\ \hline
$\hat{\phi}^{\prime}_{L}$ & $\phi^{\prime}_{L}$ & $\tilde{\phi}^{\prime}_{L}$ &  $F_{\phi^{\prime}_{L}}$ \\ \hline
$\hat{\phi}^{\prime}_{R}$ & $\phi^{\prime}_{R}$ & $\tilde{\phi}^{\prime}_{R}$ & $F_{\phi^{\prime}_{R}}$ \\ \hline
$\hat{S}$ & $S$ & $\tilde{S}$ &  $F_{S}$ \\ \hline
\end{tabular}
\end{center}
\caption{Particle content in each chiral superfield defined for the scalars defined in Eqs.(\ref{usualscalars},\ref{newscalars}).}
\label{scalar}
\end{table} 

The scalars are introduced 
at followings chiral superfields: $\hat{\phi}_{L,R},
\hat{\phi}^{\prime}_{L,R}$ and $\hat{S}$, see Tab.(\ref{scalar}). Their vaccum expectation values (vev) are
\begin{eqnarray}
\langle \phi_{L}\rangle &=& 
\frac{v_{L}}{\sqrt{2}},
\langle \phi^{\prime}_{L}\rangle =
\frac{v^{\prime}_{L}}{\sqrt{2}}, \nonumber \\
\langle \phi_{R}\rangle &=& 
\frac{v_{R}}{\sqrt{2}},
\langle \phi^{\prime}_{R}\rangle =
\frac{v^{\prime}_{R}}{\sqrt{2}}, \nonumber \\
\langle S \rangle &=& 
\frac{x}{\sqrt{2}}.
\label{vevofthemodel}
\end{eqnarray}

In the non-SUSY model LRUSM gauge group breaks to the 
Standard Model (SM) gauge group when $\phi_{R}$ acquires a vev and the SM gauge group breaks to $SU(3)_{C}\otimes U(1)_{EM}$ 
when $\phi_{L}$ acquires vev, 
then we have the following constraint
\begin{equation}
v_{R}\gg v_{L},
\label{lrusmconstraint}
\end{equation} 
due it we can choose at MSUSM 
\begin{equation}
v^{\prime}_{R} \simeq v_{R}\gg v_{L} \simeq v^{\prime}_{L},
\label{msusmconstraint}
\end{equation} 

The sleptons and squarks do not get vev.

\section{The Lagrangian}
\label{sec:lagrangian}

The supersymmetric invariant Lagrangian of the model is built with superfields given in Sec.~\ref{sec:particles}. It has the following form
\begin{equation}
{\cal L}_{MSUSM}= 
{\cal L}_{SUSY} + 
{\cal L}_{soft}, \label{l1}
\end{equation}
where, as usual, ${\cal L}_{SUSY}$ is the supersymmetric piece, while ${\cal L}_{soft}$ explicitly breaks SUSY.
Below we  write ${\cal L}_{SUSY}$ in terms of the respective superfields, while in Subsec.~\ref{subsec:soft}
we write ${\cal L}_{soft}$ in terms of the fields.

\subsection{The supersymmetric terms}
\label{subsec:susyterms}

The supersymmetric term can be divided as follows
\begin{equation}
{\cal L}_{SUSY} = 
{\cal L}_{Lepton} + 
{\cal L}_{Quarks} + 
{\cal L}_{Gauge} + 
{\cal L}_{Scalar},
\label{l2}
\end{equation}
The first term in Eq.~(\ref{l2}) is given by
\begin{eqnarray}
{\cal L}_{Lepton}&=& 
\int d^{4}\theta\; \left[\,
\hat{ \bar{L}}_{L}
e^{2[g_{L}\hat{V}_{L}+g_{BL} \left( -\frac{1}{2} \right) \hat{b}_{BL}]}\hat{L}_{L}+
\hat{ \bar{L}}_{R}
e^{2[g_{R}\hat{V}_{R}+g_{BL} \left( +\frac{1}{2} \right) \hat{b}_{BL}]}\hat{L}_{R} 
\right. \nonumber \\ 
&+& \left.
\hat{ \bar{E}}^{H}_{L} e^{2[g_{BL} \left( -\frac{2}{2} \right) \hat{V}_{BL}]}
\hat{E}^{H}_{L}+
\hat{ \bar{E}}^{H}_{R} e^{2[g_{BL} \left( +\frac{2}{2} \right) \hat{V}_{BL}]}
\hat{E}^{H}_{R}+
\hat{ \bar{N}}^{H}_{L} \hat{N}^{H}_{L}+
\hat{ \bar{N}}^{H}_{R} \hat{N}^{H}_{R}
\right]. \nonumber \\
\label{l3charg}
\end{eqnarray}
In the expressions above we have used $\hat{V}_{L}=T^{i}\hat{V}^{i}_{L}$ and 
$\hat{V}_{R}=T^{i}\hat{V}^{i}_{R}$
where $T^{i}=\sigma^{i}/2$ (with $i=1,2,3$) are the
generators of $SU(2)_{L}$ 
and $SU(2)_{R}$ while $g_{BL}$ are the gauge constant constants of the $U(1)_{B-L}$, respectively, as showed in Table~\ref{table1}.

The second term in Eq.(\ref{l2}) is written as
\begin{eqnarray}
{\cal L}_{Quarks}&=& \int d^{4}\theta\; \left[\,
\hat{ \bar{Q}}_{L}
e^{2[g_{s}\hat{G}+
g_{L}\hat{V}_{L}+g_{BL} \left( +\frac{1}{6} \right) \hat{b}_{BL}]}\hat{Q}_{L}+
\hat{ \bar{Q}}_{R}
e^{2[g_{s}\hat{\bar{G}}+
g_{R}\hat{V}_{R}+g_{BL} \left( -\frac{1}{6} \right) \hat{b}_{BL}]}\hat{Q}_{R} 
\right. \nonumber \\ 
&+& \left.
\hat{ \bar{U}}^{H}_{L} e^{2[g_{s}\hat{G}+
g_{BL} \left( +\frac{4}{6} \right) \hat{V}_{BL}]}
\hat{U}^{H}_{L}+
\hat{ \bar{U}}^{H}_{R} e^{2[g_{s}\hat{\bar{G}}+
g_{BL} \left( -\frac{4}{6} \right) \hat{V}_{BL}]}
\hat{U}^{H}_{R}
\right. \nonumber \\ 
&+& \left.
\hat{ \bar{D}}^{H}_{L}
e^{2[g_{s}\hat{G}+
g_{BL} \left( -\frac{2}{6} \right) \hat{V}_{BL}]} \hat{D}^{H}_{L}+
\hat{ \bar{D}}^{H}_{R}
e^{2[g_{s}\hat{\bar{G}}+
g_{BL} \left( +\frac{2}{6} \right) \hat{V}_{BL}]} \hat{D}^{H}_{R}
\right]
\label{quarks}
\end{eqnarray}
where $\hat{G}=T^{a}\hat{G}^{a}$, 
$\hat{\bar{G}}=T^{a*}\hat{G}^{a}$ and $T^{a}=(\lambda^{a}/2)$ 
(with $a=1,2, \ldots ,8$) 
are the Gell-Mann matrixes the generators of $SU(3)_{C}$.

The gauge part is given by
\begin{eqnarray}
{\cal L}_{Gauge} &=& \frac{1}{4} \int d^{2}\theta\; \left[ \sum_{a=1}^{8}W^{a}_{c}W^{a}_{c}
+ \sum_{i=1}^{3} 
W^{i}_{L}W^{i}_{L}+
\sum_{i=1}^{3} 
W^{i}_{R}W^{i}_{R}
+W^{BL}W^{BL} \right]
+ H.c. \,\ , \nonumber \\
\label{l5}
\end{eqnarray}
where the strength fields are defined as
\begin{eqnarray}
W^{a}_{\alpha c}&=&- \frac{1}{8g_{s}} \bar{D} \bar{D} e^{-2g_{s} \hat{G}^{a}}
D_{\alpha} e^{2g_s \hat{G}^{a}},  
\nonumber \\
W^{i}_{\alpha L}&=&- \frac{1}{8g_{L}} \bar{D} \bar{D} e^{-2g_{L} \hat{V}^{i}_{L}} D_{\alpha}
e^{2g_{L} \hat{V}^{i}_{L}},  \nonumber \\
W^{i}_{\alpha R}&=&- \frac{1}{8g_{R}} \bar{D} \bar{D} e^{-2g_{R} \hat{V}^{i}_{R}} D_{\alpha}
e^{2g_{R} \hat{V}^{i}_{R}},  
\nonumber \\
W^{BL}_{\alpha}&=&- \frac{1}{4} \bar{D} \bar{D} D_{\alpha} \hat{V}^{BL} \,\ ,
\label{l6}
\end{eqnarray}
where $D_{\alpha}$ is the covariant derivative and it is given by \cite{Rodriguez:2019mwf}:
\begin{eqnarray}
D_{\alpha}(y,\theta,\bar{\theta}) &=& \frac{\partial}{\partial \theta^{\alpha}}+2i 
\sigma^{m}_{\alpha \dot{\alpha}}\bar{\theta}^{\dot{\alpha}}\frac{\partial}{\partial y^{m}} \,\ \nonumber \\ 
\bar{D}_{\dot{\alpha}}(y,\theta,\bar{\theta})&=&- \frac{\partial}{\partial \bar{\theta}^{\dot{\alpha}}} \,\ .
\label{The Non-Abelian Fieldstrength prop 4}
\end{eqnarray}

Finally, the scalar part in Eq.(\ref{l2}) is
\begin{eqnarray}
{\cal L}_{Scalar} &=& \int d^{4}\theta\;\left[\,
\hat{ \overline{\phi_{L}}}
e^{2[g_{L}\hat{V}_{L}+g_{BL} \left( -\frac{1}{2} \right) \hat{b}_{BL}]}\hat{\phi}_{L}+
\hat{ \overline{\phi_{R}}}
e^{2[g_{R}\hat{V}_{R}+g_{BL} \left( -\frac{1}{2} \right) \hat{b}_{BL}]}\hat{\phi}_{R}
\right. \nonumber \\
&+& \left.
\overline{\phi^{\prime}_{L}}
e^{2[g_{L}\hat{V}_{L}+g_{BL} \left( +\frac{1}{2} \right) \hat{b}_{BL}]}
\hat{\phi}^{\prime}_{L}+
\hat{ \overline{\phi^{\prime}_{R}}}
e^{2[g_{R}\hat{V}_{R}+g_{BL} \left( \frac{1}{2} \right) \hat{b}_{BL}]}
\hat{\phi}^{\prime}_{R}+
\hat{\bar{S}}\hat{S}
\right]  \nonumber \\
&+& \left(\int d^2\theta\, W+ H.c.
\right),  \label{l7}
\end{eqnarray}
where $W$ is the superpotential, which we discuss in the next subsection.

\subsection{The superpotential}
\label{subsec:superp}

The superpotential of the model is given by
\begin{equation}
W=W_{1}+W_{2}+W_{3},  \label{sp1m1}
\end{equation}
with 
\begin{eqnarray}
W_{1}&=&\zeta_{S}\hat{S}+
\zeta_{N^{H}_{L}}
\hat{N}^{H}_{L}+
\zeta_{N^{H}_{R}}
\hat{N}^{H}_{R}, 
\nonumber \\
W_{2}&=&\mu_{L}\left( 
\hat{\phi}_{L}
\hat{\phi}^{\prime}_{L}
\right) + 
\mu_{R}\left( 
\hat{\phi}_{R}
\hat{\phi}^{\prime}_{R}
\right) + 
\frac{\mu_{S}}{2}
\hat{S}\hat{S}+ 
\frac{\mu_{N^{H}_{L}}}{2}
\hat{N}^{H}_{L} \hat{N}^{H}_{L} + 
\frac{\mu_{N^{H}_{R}}}{2} \hat{N}^{H}_{R} \hat{N}^{H}_{R} +
\mu_{LR}\hat{N}^{H}_{L}
\hat{N}^{H}_{R} 
\nonumber \\ 
&+&
\mu_{SL}\hat{S}
\hat{N}^{H}_{L}+
\mu_{SR}\hat{S}
\hat{N}^{H}_{R}, 
\nonumber \\
W_{3}&=&
\frac{\kappa_{S}}{3}
\hat{S}\hat{S}\hat{S}+
\lambda_{L}\left( 
\hat{\phi}_{L}
\hat{\phi}^{\prime}_{L}
\right) \hat{S} + 
\lambda_{R}\left( 
\hat{\phi}_{R}
\hat{\phi}^{\prime}_{R}
\right) \hat{S}+
\frac{\kappa^{1}}{3}
\hat{N}^{H}_{L} 
\hat{N}^{H}_{L} \hat{N}^{H}_{L}+
\frac{\kappa^{2}}{3}
\hat{N}^{H}_{R} 
\hat{N}^{H}_{R} \hat{N}^{H}_{R} 
\nonumber \\ 
&+&
\kappa^{3}
\hat{N}^{H}_{L} 
\hat{N}^{H}_{L} \hat{N}^{H}_{R}+
\kappa^{4}
\hat{N}^{H}_{L} 
\hat{N}^{H}_{L} 
\hat{S} +
\kappa^{5}
\hat{N}^{H}_{R} 
\hat{N}^{H}_{R} \hat{N}^{H}_{L} +
\kappa^{6}
\hat{N}^{H}_{R} 
\hat{N}^{H}_{R} 
\hat{S} 
\nonumber \\
&+&
\lambda^{L}_{NHL}\left( 
\hat{\phi}_{L}
\hat{\phi}^{\prime}_{L}
\right) \hat{N}^{H}_{L} + 
\lambda^{R}_{NHL}\left( 
\hat{\phi}_{R}
\hat{\phi}^{\prime}_{R}
\right) \hat{N}^{H}_{L}+
\lambda^{L}_{NHR}\left( 
\hat{\phi}_{L}
\hat{\phi}^{\prime}_{L}
\right) \hat{N}^{H}_{R} + 
\lambda^{R}_{NHR}\left( 
\hat{\phi}_{R}
\hat{\phi}^{\prime}_{R}
\right) \hat{N}^{H}_{R} 
\nonumber \\
&+&  
G_{u}\hat{U}^{H}_{L}
\hat{U}^{H}_{R}\hat{S}+
G_{u1}\hat{U}^{H}_{L}
\hat{U}^{H}_{R}
\hat{N}^{H}_{L}+
G_{u2}\hat{U}^{H}_{L}
\hat{U}^{H}_{R}
\hat{N}^{H}_{R}+
G_{d}\hat{D}^{H}_{L}
\hat{D}^{H}_{R}\hat{S}+
G_{d1}\hat{D}^{H}_{L}
\hat{D}^{H}_{R}
\hat{N}^{H}_{L} 
\nonumber \\
&+&
G_{d2}\hat{D}^{H}_{L}
\hat{D}^{H}_{R}
\hat{N}^{H}_{R}+ 
G_{e}\hat{E}^{H}_{L}
\hat{E}^{H}_{R}\hat{S}+
G_{e1}\hat{E}^{H}_{L}
\hat{E}^{H}_{R}
\hat{N}^{H}_{L}+
G_{e2}\hat{E}^{H}_{L}
\hat{E}^{H}_{R}
\hat{N}^{H}_{R} 
\nonumber \\
&+&
Y_{uL}\left( \hat{Q}_{L} 
\hat{\phi}^{\prime}_{L}
\right) \hat{U}^{H}_{R}+ 
Y_{uR}\left( \hat{Q}_{R} 
\hat{\phi}_{R}
\right) \hat{U}^{H}_{L}+
Y_{dL}\left( \hat{Q}_{L} 
\hat{\phi}_{L}
\right) \hat{D}^{H}_{R}+
Y_{dR}\left( \hat{Q}_{R} 
\hat{\phi}^{\prime}_{R}
\right) \hat{D}^{H}_{L} 
\nonumber \\ 
&+&
Y_{eL}\left( \hat{L}_{L} 
\hat{\phi}_{L}
\right) \hat{E}^{H}_{R}+
Y_{eR}\left( \hat{L}_{R} 
\hat{\phi}^{\prime}_{R}
\right) \hat{E}^{H}_{L}+
Y_{\nu L}\left( \hat{L}_{L} 
\hat{\phi}^{\prime}_{L}
\right) \hat{N}^{H}_{R}+
Y_{\nu R}\left( \hat{L}_{R} 
\hat{\phi}_{R}
\right) \hat{N}^{H}_{L},
\nonumber \\
\label{sp3m1}
\end{eqnarray}
where we have defined 
$\left( 
\hat{\phi}_{L}
\hat{\phi}^{\prime}_{L}
\right) \equiv \epsilon_{\alpha \beta} \hat{\phi}^{\alpha}_{L}
\hat{\phi}^{\prime \beta}_{L}$ as done in the 
Minimal Supersymmetric Standard Model (MSSM) \cite{Rodriguez:2019mwf}.

We defined at Tab.(\ref{allrpqchargesinMSUSM}) the $R$-charges of the superfields in the MSUSM.  
\begin{table}[h]
\begin{center}
\begin{tabular}{|c|c|}
\hline  
$\mbox{ Superfield}$ & $R$-charge \\
\hline
$\hat{L}_{L,R}$ & $n_{L_{L,R}}=+ \left( \frac{1}{2} \right)$ \\ 
\hline
$\hat{Q}_{L,R}$ & $n_{Q_{L,R}}=+ \left( \frac{1}{2} \right)$ \\ 
\hline
$\hat{f}^{H}_{L,R}$ & $n_{{f}^{H}_{L,R}}=+ \left( \frac{1}{2} \right)$ \\ 
\hline
$\hat{\phi}_{L,R}$ & $n_{\phi_{L,R}}=+ 1$ \\ 
\hline
$\hat{\phi}^{\prime}_{L,R}$ & $n_{\phi^{\prime}_{L,R}}=+ 1$ \\ 
\hline
$\hat{S}$ & $n_{S}=+ 1$ \\ 
\hline
\end{tabular}
\end{center}
\caption{\small $R$-charge assignment to all superfields in the MSUSM and 
$f=U,D,E$ and $N$.}
\label{allrpqchargesinMSUSM}
\end{table}

The terms of our superpotential that conserves our $R$-parity, 
defined at Tab.(\ref{allrpqchargesinMSUSM}), are given by
\begin{eqnarray}
W_{RC}&=&W_{2RC}+W_{3RC}, 
\nonumber \\  
W_{2RC}&=&\mu_{L}\left( 
\hat{\phi}_{L}
\hat{\phi}^{\prime}_{L}
\right) + 
\mu_{R}\left( 
\hat{\phi}_{R}
\hat{\phi}^{\prime}_{R}
\right) + 
\frac{\mu_{S}}{2}
\hat{S} 
\hat{S}, \nonumber \\
W_{3RC}&=&G_{u}\hat{U}^{H}_{L}
\hat{U}^{H}_{R}\hat{S}+
G_{d}\hat{D}^{H}_{L}
\hat{D}^{H}_{R}\hat{S} +
G_{e}\hat{E}^{H}_{L}
\hat{E}^{H}_{R}\hat{S}+
G_{1} \hat{N}^{H}_{L} 
\hat{N}^{H}_{L} \hat{S}+
G_{2}\hat{N}^{H}_{L}
\hat{N}^{H}_{R}\hat{S} 
\nonumber \\
&+&
G_{3}\hat{N}^{H}_{R} 
\hat{N}^{H}_{R} \hat{S}+
Y_{uL}\left( \hat{Q}_{L} 
\hat{\phi}^{\prime}_{L}
\right) \hat{U}^{H}_{R} +
Y_{uR}\left( \hat{Q}_{R} 
\hat{\phi}_{R}
\right) \hat{U}^{H}_{L}+
Y_{dL}\left( \hat{Q}_{L} 
\hat{\phi}_{L}
\right) \hat{D}^{H}_{R} 
\nonumber \\ 
&+&
Y_{dR}\left( \hat{Q}_{R} 
\hat{\phi}^{\prime}_{R}
\right) \hat{D}^{H}_{L}+
Y_{eL}\left( \hat{L}_{L} 
\hat{\phi}_{L}
\right) \hat{E}^{H}_{R} +
Y_{eR}\left( \hat{L}_{R} 
\hat{\phi}^{\prime}_{R}
\right) \hat{E}^{H}_{L}+
Y_{\nu L}\left( \hat{L}_{L} 
\hat{\phi}^{\prime}_{L}
\right) \hat{N}^{H}_{R} 
\nonumber \\ 
&+&
Y_{\nu R}\left( \hat{L}_{R} 
\hat{\phi}_{R}
\right) \hat{N}^{H}_{L}.
\nonumber \\
\label{sp3m1}
\end{eqnarray}
The coefficients $\mu_{L}$,$\mu_{R}$ and $\mu_{S}$ have mass dimension, while all the 
coeffients in $W_{3RC}$ are 
dimensionless. Those coefficients at $W_{2RC},W_{3RC}$ are, in principle, complex numbers 
\cite{Rodriguez:2019mwf}.

\subsection{Soft terms}
\label{subsec:soft}

They depend on the model under consideration and in our case they  can be written  as
\begin{eqnarray}
{\cal L}_{soft} &=& {\cal L}_{SMT} +{\cal L}_{GMT}+{\cal L}_{INT}, \nonumber \\
\label{SoftSUSYm1}
\end{eqnarray}
where the scalar mass term ${\cal L}_{SMT}$, is given by
\begin{eqnarray}
{\cal L}_{SMT} &=&- \left(
M^{2}_{L_{L}}
\vert \tilde{L}_{L} \vert^{2}+ 
M^{2}_{L_{R}}
\vert \tilde{L}_{R}
\vert^{2}+
M^{2}_{E^{H}_{L}}
\vert \tilde{E}^{H}_{L}
\vert^{2}+ 
M^{2}_{E^{H}_{R}}
\vert \tilde{E}^{H}_{R}
\vert^{2}+
M^{2}_{N^{H}_{L}}
\vert \tilde{N}^{H}_{L}
\vert^{2} 
\right. \nonumber \\
&+& \left.
M^{2}_{N^{H}_{R}}
\vert \tilde{N}^{H}_{R}
\vert^{2}
+  
M^{2}_{Q_{L}}
\vert \tilde{Q}_{L}
\vert^{2}+ 
M^{2}_{Q_{R}}
\vert \tilde{Q}_{R}
\vert^{2}+
M^{2}_{D^{H}_{L}}
\vert \tilde{D}^{H}_{L}
\vert^{2}+ 
M^{2}_{D^{H}_{R}}
\vert \tilde{D}^{H}_{R}
\vert^{2}+
M^{2}_{U^{H}_{L}}
\vert \tilde{U}^{H}_{L}
\vert^{2}
\right. \nonumber \\
&+& \left. 
M^{2}_{U^{H}_{R}}
\vert \tilde{U}^{H}_{R}
\vert^{2} 
+
\vert \phi^{\prime}_{R} \vert^{2}+ 
\right) 
+ \left[
\beta^{2}_{\phi_{L}} \left( 
\phi_{L} \phi^{\prime}_{L} 
\right) 
+
\beta^{2}_{\phi_{R}} \left( 
\phi_{R} \phi^{\prime}_{R} 
\right) +
\beta^{2}_{S} \left( 
S \right)^{2} + H.c. \right],
\label{smtsoft}
\end{eqnarray}
where all the coefficients having the dimension of squared mass \cite{Rodriguez:2019mwf}.

The gaugino mass term 
${\cal L}_{GMT}$ is defined as
\begin{eqnarray}
{\cal L}^{MSSM}_{GMT} &=&-\frac{1}{2} \left(\,M_{\tilde{g}}\; \sum_{a=1}^{8}
\tilde{g}^{a} \tilde{g}^{a} +
M_{\tilde{V}_{L}}\; \sum_{i=1}^{3}\;\tilde{V}^{i}_{L} \tilde{V}^{i}_{L}+
M_{\tilde{V}_{R}}\; \sum_{i=1}^{3}\;\tilde{V}^{i}_{R} \tilde{V}^{i}_{R} +
M_{\tilde{b}_{BL}}\; \tilde{V}^{BL} \tilde{V}{BL} \,\right) 
\nonumber \\ &+&
H.c., 
\end{eqnarray}
and the last term ${\cal L}_{INT}$ is
\begin{eqnarray}
{\cal L}_{INT} &=&A^{u}G_{u}
\tilde{U}^{H}_{L}
\tilde{U}^{H}_{R}S+
A^{d}G_{d}
\tilde{D}^{H}_{L}
\tilde{D}^{H}_{R}S +
A^{e}G_{e}
\tilde{E}^{H}_{L}
\tilde{E}^{H}_{R}S+
A^{1}G_{1}
\tilde{N}^{H}_{L}
\tilde{N}^{H}_{L}S 
\nonumber \\
&+&
A^{2}G_{2}
\tilde{N}^{H}_{L}
\tilde{N}^{H}_{R}S+
A^{3}G_{3}
\tilde{N}^{H}_{R}
\tilde{N}^{H}_{R}S+
B^{uL}Y_{uL}
\left( \tilde{Q}_{L} 
\phi^{\prime}_{L}
\right) \tilde{U}^{H}_{R} +
B^{uR}Y_{uR}
\left( \tilde{Q}_{R} 
\phi_{R}
\right) \tilde{U}^{H}_{L}
\nonumber \\
&+&
B^{dL}Y_{dL}
\left( \tilde{Q}_{L} 
\phi_{L}
\right) \tilde{D}^{H}_{R}+
B^{dR}Y_{dR}
\left( \tilde{Q}_{R} 
\phi^{\prime}_{R}
\right) \tilde{D}^{H}_{L}+
B^{eL}Y_{eL}\left( \tilde{L}_{L} 
\phi_{L}
\right) \tilde{E}^{H}_{R} 
\nonumber \\
&+&
B^{eR}Y_{eR}
\left( \tilde{L}_{R} 
\phi^{\prime}_{R}
\right) \tilde{E}^{H}_{L}+
B^{\nu L}Y_{\nu L}
\left( \tilde{L}_{L} 
\phi^{\prime}_{L}
\right) \tilde{N}^{H}_{R}+
B^{\nu R}Y_{\nu R}
\left( \tilde{L}_{R} 
\phi_{R}
\right) \tilde{N}^{H}_{L}. 
\label{softint}
\end{eqnarray}

\section{Fermion Masses}

We will calculate the masses 
to all the usual fermions of this model. Their masses 
can be get from our superpotential, see 
Eq.(\ref{sp3m1}). Therefore, 
the masses of the fermions came from 
\begin{eqnarray}
&&G_{u}U^{H}_{L}U^{H}_{R}S+ 
Y_{uL}\left( Q_{L} 
\phi^{\prime}_{L}
\right) U^{H}_{R} +
Y_{uR}\left( Q_{R}\phi_{R}
\right) U^{H}_{L}+ 
G_{d}D^{H}_{L}D^{H}_{R}S+
Y_{dL}\left( Q_{L}\phi_{L}
\right) D^{H}_{R}
\nonumber \\ 
&+&  
Y_{dR}\left( Q_{R} 
\phi^{\prime}_{R}
\right) D^{H}_{L}+
G_{e}E^{H}_{L}E^{H}_{R}S+
Y_{eL}\left( L_{L}\phi_{L}
\right) E^{H}_{R} +
Y_{eR}\left( L_{R} 
\phi^{\prime}_{R}
\right) E^{H}_{L}+
G_{1}N^{H}_{L}N^{H}_{L}S
\nonumber \\
&+&
G_{2}N^{H}_{L}N^{H}_{R}S+ 
G_{3}N^{H}_{R}N^{H}_{R}S+
Y_{\nu L}\left( L_{L} 
\phi^{\prime}_{L}
\right) N^{H}_{R}+ 
Y_{\nu R}\left( L_{R} \phi_{R}
\right) N^{H}_{L}+ H.c.
\end{eqnarray}

Therefore, the ``down" quark sector ($d,s$ and $b$ quarks) as well as the
$e,\mu$ and $\tau$ will have masses proportional to the vacuum expectation
values $v_{L},v^{\prime}_{R}$, whereas the ``up" sector and for neutrinos, we will have masses proportional to
$v_{R},v^{\prime}_{L}$.

After symmetry breaking, we 
get the following mass matrices for the usual fermions
\begin{eqnarray}
{\cal M}_{uU}&=&
\frac{1}{\sqrt{2}} \left( 
\begin{array}{cc} 
0 & Y_{uL}v^{\prime}_{L} \\ 
Y_{uR}v_{R} & \xi_{u} 
\end{array} 
\right), \,\
{\cal M}_{dD}=
\frac{1}{\sqrt{2}} \left( 
\begin{array}{cc} 
0 & Y_{dL}v_{L} \\ 
Y_{dR}v^{\prime}_{R} & \xi_{d} 
\end{array} 
\right), \nonumber \\
{\cal M}_{eE}&=&
\frac{1}{\sqrt{2}} \left( 
\begin{array}{cc} 
0 & Y_{eL}v_{L} \\ 
Y_{eR}v^{\prime}_{R} & \xi_{e} 
\end{array} 
\right), \,\
{\cal M}_{\nu N}=
\frac{1}{\sqrt{2}} \left( 
\begin{array}{cc} 
0 & Y_{\nu L}v^{\prime}_{L} \\ 
Y_{\nu R}v_{R} & \xi_{\nu} 
\end{array} 
\right), \nonumber \\
\end{eqnarray}
where 
\begin{eqnarray}
\xi_{u}&=&G_{u}x, 
\,\
\xi_{d}=G_{d}x, 
\nonumber \\
\xi_{e}&=&G_{e}x, 
\,\
\xi_{\nu}=G_{\nu}x,
\label{usmechanism}
\end{eqnarray}
those mass matrices are identical as presented at 
\cite{Davidson:1987mh} and $x$ is the vev of the scalar Singlet, see 
Eq.(\ref{vevofthemodel}). Their mass eigenstates, 
mixing angle and the left and right orthogonal 
transformations can be found 
at \cite{Hati:2018tge}.

We have introduced only one fermion familly as done at \cite{Davidson:1987mh}, but in the case of Supersymmetric model 
it is not a problem because we can generate masses for the quarks and 
charged leptons, at second 
and third families, throught a 
radiative mechanism \cite{banks,ma,cmmc,cmmc1}. 

Therefore we can introduce the second and thir fermion famillies in such way as they do not couple to our usual Scalars bur 
they interact with the usual Gauge Bosons. We can also 
to explain any mixing parameter given by Cabibbo-Kobayashi-Maskawa (CKM) matrix, in the quark sector \cite{cmmc,cmmc1}. 

We can also explain the masses of two neutrinos and 
the Pontecorvo-Maki-Nakagawa-Sakata (PMNS) matrix 
in similar ways as done at \cite{global}. 

We can also consider the two-generation families, 
it was presented at \cite{Davidson:1987tr}. The three families case, requires a more detailed quantitative analysis will be studies later.

We can define the following Dirac four-components spinos
\begin{eqnarray}
{\cal E}_{L}&=&\left( 
\begin{array}{c}
e^{-} \\ 
\bar{E}^{H}_{R}
\end{array}
\right) \,\
E_{R}=\left( 
\begin{array}{c}
e^{+} \\ 
\bar{E}^{H}_{L}
\end{array}
\right) \nonumber \\
{\cal N}_{L}&=&\left( 
\begin{array}{c}
\nu_{e} \\ 
\bar{N}^{H}_{R}
\end{array}
\right) \,\
N_{R}=\left( 
\begin{array}{c}
\bar{\nu}_{e} \\ 
\bar{N}^{H}_{L}
\end{array}
\right) \nonumber \\
{\cal U}_{L}&=&\left( 
\begin{array}{c}
u \\ 
\bar{U}^{H}_{R}
\end{array}
\right) \,\
U_{R}=\left( 
\begin{array}{c}
u^{c} \\ 
\bar{U}^{H}_{L}
\end{array}
\right) \nonumber \\
{\cal D}_{L}&=&\left( 
\begin{array}{c}
d \\ 
\bar{D}^{H}_{R}
\end{array}
\right) \,\
D_{R}=\left( 
\begin{array}{c}
d^{c} \\ 
\bar{D}^{H}_{L}
\end{array}
\right).
\label{4Diracespinor}
\end{eqnarray}

\section{Gauge Bosons Masses}

We are going to study the gauge boson 
sector.
\begin{eqnarray}
\left( {\cal D}_{m}
\phi_{L}
\right)^{\dagger}
\left( {\cal D}_{m}
\phi_{L}\right) + 
\left( {\cal D}_{m}
\phi_{R}\right)^{\dagger}
\left( {\cal D}_{m}
\phi_{R}\right) +
\left( {\cal D}_{m}
\phi^{\prime}_{L}
\right)^{\dagger}
\left( {\cal D}_{m}
\phi^{\prime}_{L}\right) + 
\left( {\cal D}_{m}
\phi^{\prime}_{R}
\right)^{\dagger}
\left( {\cal D}_{m}
\phi^{\prime}_{R}\right) ,
\label{originmassgaugebosons}
\end{eqnarray}
where ${\cal D}_{m}$ is covariant derivates of the MSUSM given by:
\begin{eqnarray}
{\cal D}_{m}\phi_{L}&=& \left( \partial_{m}+ \imath 
\frac{g_{L}}{2} 
\vec{\sigma}\cdot \vec{V}^{L}_{m}- \imath 
\frac{g_{BL}}{2}V^{BL}_{m}
\right) \phi_{L} , \nonumber \\
{\cal D}_{m}\phi_{R}&=& \left( \partial_{m}+ \imath 
\frac{g_{R}}{2} 
\vec{\sigma}\cdot \vec{V}^{R}_{m}- \imath 
\frac{g_{BL}}{2}V^{BL}_{m}
\right) \phi_{R} , \nonumber \\
{\cal D}_{m}
\phi^{\prime}_{L}&=& \left( \partial_{m}+ \imath 
\frac{g_{L}}{2} 
\vec{\sigma}\cdot \vec{V}^{L}_{m}- \imath 
\frac{g_{BL}}{2}V^{BL}_{m}
\right) \phi^{\prime}_{L} , \nonumber \\
{\cal D}_{m}
\phi^{\prime}_{R}&=& \left( \partial_{m}+ \imath 
\frac{g_{R}}{2} 
\vec{\sigma}\cdot \vec{V}^{R}_{m}- \imath 
\frac{g_{BL}}{2}V^{BL}_{m}
\right) \phi^{\prime}_{R}.
\end{eqnarray}
We can calculate from those expression the masses of the charged bosons and also the masses of the neutrals bosons.

\subsection{Charged Gauge Bosons}
After some simple calculation, we get the following expression to the masses of the charged ones 
\begin{equation}
\left( W^{\pm}_{L} \right)_{m}= 
\frac{1}{\sqrt{2}}\left[ 
\left( V^{1}_{L}\right)_{m} \mp \imath 
\left( V^{2}_{L}\right)_{m}\right], 
\nonumber \\
\left( W^{\pm}_{R} \right)_{m}= 
\frac{1}{\sqrt{2}}\left[ 
\left( V^{1}_{R}\right)_{m} \mp \imath 
\left( V^{2}_{R}\right)_{m}\right],
\label{wdef}
\end{equation} 
get the following mass
\begin{eqnarray}
M^{2}_{W^{\pm}_{L}}&=& \frac{g^{2}_{L}}{4}\left( v_{L}^{2}+
v_{L}^{\prime \,\ 2}
\right)=
\frac{g^{2}_{L}v_{L}^{2}}{4}
\left( 1+ \tan^{2}\beta
\right)=
\frac{g^{2}_{L}v_{L}^{2}}{4} 
\sec^{2}\beta , \nonumber \\
M^{2}_{W^{\pm}_{R}}&=& \frac{g^{2}_{R}}{4}\left(
v_{R}^{2}+
v_{R}^{\prime \,\ 2}
\right)=
\frac{g^{2}_{R}v_{R}^{2}}{4}
\left( 1+ \tan^{2}\alpha
\right)=
\frac{g^{2}_{R}v_{R}^{2}}{4} 
\sec^{2}\alpha ,
\label{chargedgaugebosonsmasses}
\end{eqnarray}
where
\begin{eqnarray}
\tan \beta \equiv 
\frac{v^{\prime}_{L}}{v_{L}}, 
\,\
\tan \alpha \equiv 
\frac{v^{\prime}_{R}}{v_{R}},
\label{defbetaalpha} 
\end{eqnarray} 
and in similar way as in the MSSM \cite{Rodriguez:2019mwf}, we can say that both $\beta$ and $\alpha$-parameters are 
free parameter on the theory and due Eq.(\ref{msusmconstraint}) we can realize
\begin{equation}
M_{W^{\pm}_{L}}\ll M_{W^{\pm}_{R}},
\label{mwlbigmwr}
\end{equation}
as our first result.

\begin{figure}[ht]
\begin{center}
\vglue -0.009cm
\mbox{\epsfig{file=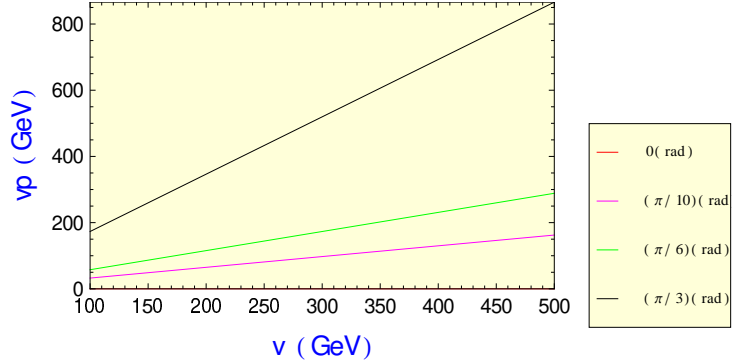,width=0.7\textwidth,
angle=0}}       
\end{center}
\caption{We show the values for $vp \equiv
v^{\prime}_{L}$ as function of $v\equiv v_{L}$, using the first relation defined at 
Eq.(\ref{defbetaalpha}), for $\beta = 0, ( \pi /10), 
( \pi /6), ( \pi /3)$ rad as indicated at the box on right.}
\label{fig01}
\end{figure}

\begin{figure}[ht]
\begin{center}
\vglue -0.009cm
\mbox{\epsfig{file=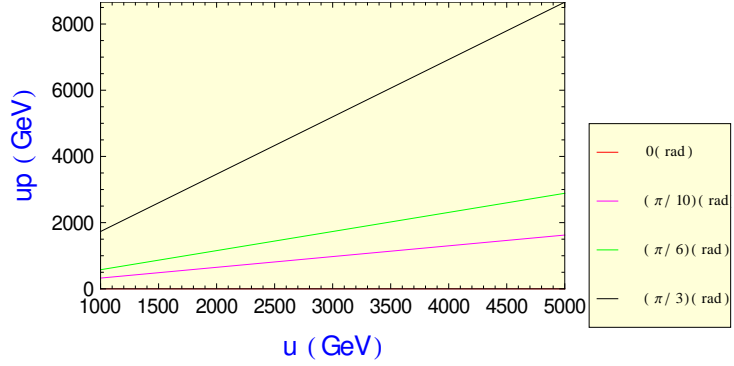,width=0.7\textwidth,
angle=0}}       
\end{center}
\caption{We show the values for $up \equiv
v^{\prime}_{R}$ as function of $u\equiv v_{R}$, using the second relation defined at 
Eq.(\ref{defbetaalpha}), for $\alpha = 0, ( \pi /10), 
( \pi /6), ( \pi /3)$ rad as indicated at the box on right.}
\label{fig02}
\end{figure}

The possible values for $v^{\prime},u^{\prime}$ in terms of $v,u$ are shown at 
Figs.(\ref{fig01},\ref{fig02}). On the first 
figures we have choose 
$v_{L} \sim {\cal O}(100 \,\ {\mbox GeV})$  and we got 
$v^{\prime}_{L} \sim {\cal O}(100 \,\ {\mbox GeV})$ and 
on the second figure we have 
used $v_{R} \sim {\cal O}(1000 \,\ {\mbox GeV})$ and as result 
$v^{\prime}_{R} \sim {\cal O}(1000 \,\ {\mbox GeV})$. Those results satisfy Eq.(\ref{msusmconstraint}).

In our analyses we will consider
\begin{equation}
g= \sqrt{\frac{8G_{F}M^{2}_{W}}{\sqrt{2}}}=0.653,
\label{gSM}
\end{equation}
the Standard Model gauge constant is the same as, it means $g_{L}\equiv g$ defined at Eq.(\ref{gSM}). The experimental value of $W^{\pm}$-mass is given by:
\begin{equation}
M_{W^{\pm}}=80.379 \pm 0.012 \,\  
{\mbox GeV},
\label{WLmassconstraints}
\end{equation} 

\begin{figure}[ht]
\begin{center}
\vglue -0.009cm
\mbox{\epsfig{file=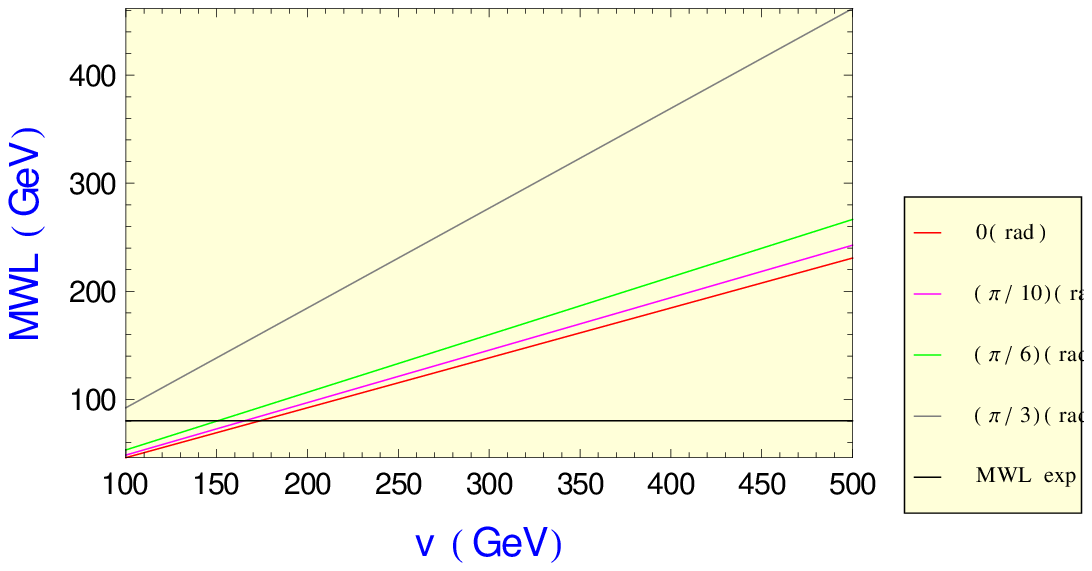,width=0.7\textwidth,
angle=0}}       
\end{center}
\caption{The masses of $W^{\pm}_{L}$ to several values of the $\beta$ parameter, with the same description give at 
Fig.(\ref{fig01}). The black line means the experimental 
values of 
$M_{W^{\pm}} \equiv M_{W^{\pm}_{L}}$, 
see Eq.(\ref{WLmassconstraints}).}
\label{fig1}
\end{figure}

Using Eqs.(\ref{chargedgaugebosonsmasses},\ref{gSM}), we can get the following plot shown 
at Fig.(\ref{fig1}), and it is the same as presented at \cite{Rodriguez:2019mwf}, and we see we can get the $W^{\pm}$-mass values for all $\beta$-parameter and $140 \leq v \leq 180$ (GeV), the exception is for $\beta = ( \pi /3)$ rad.  

The experimental constraints for the new charged gauge boson is \cite{Zyla:2020zbs}: 
\begin{equation}
M_{W^{\pm}_{R}}>2.7 \,\ 
{\mbox TeV},
\label{WRmassconstraints}
\end{equation}
in this model we have three possibilities for $g_{R}$:  \begin{itemize}
\item[1] $g_{R}>g_{L}$
\begin{figure}[ht]
\begin{center}
\vglue -0.009cm
\mbox{\epsfig{file=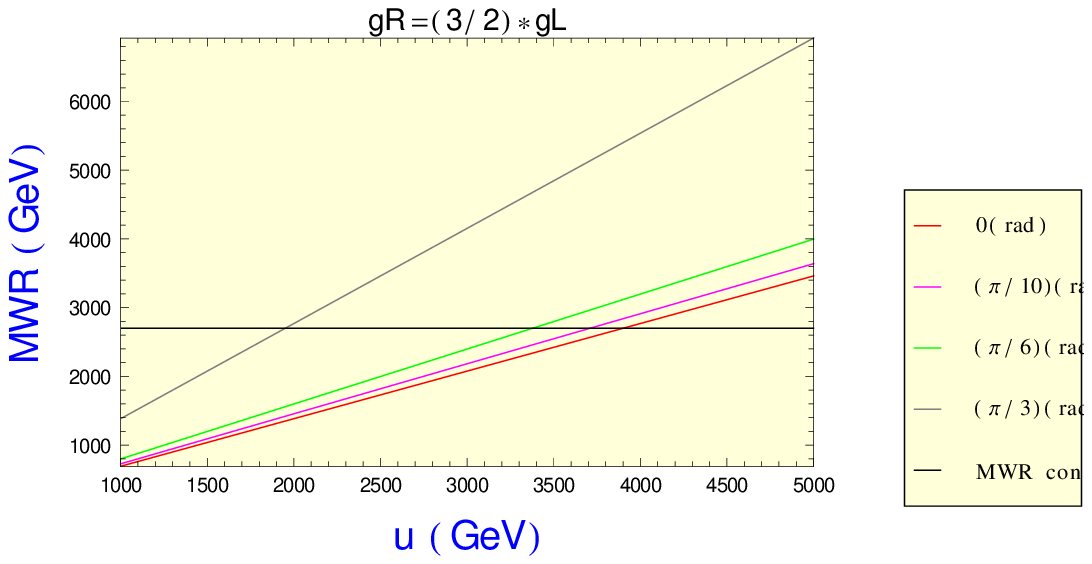,width=0.7\textwidth,
angle=0}} 
\end{center}
\caption{The masses of $W^{\pm}_{R}$ to several values of the 
$\alpha$ parameter, with the same description give at 
Fig.(\ref{fig02}). The experimental constraints of $M_{W^{\pm}_{R}}$ is 
defined at Eq.(\ref{WRmassconstraints}).}
\label{fig2}
\end{figure}
\item[2] $g_{R}=g_{L}$
\begin{figure}[ht]
\begin{center}
\vglue -0.009cm
\mbox{\epsfig{file=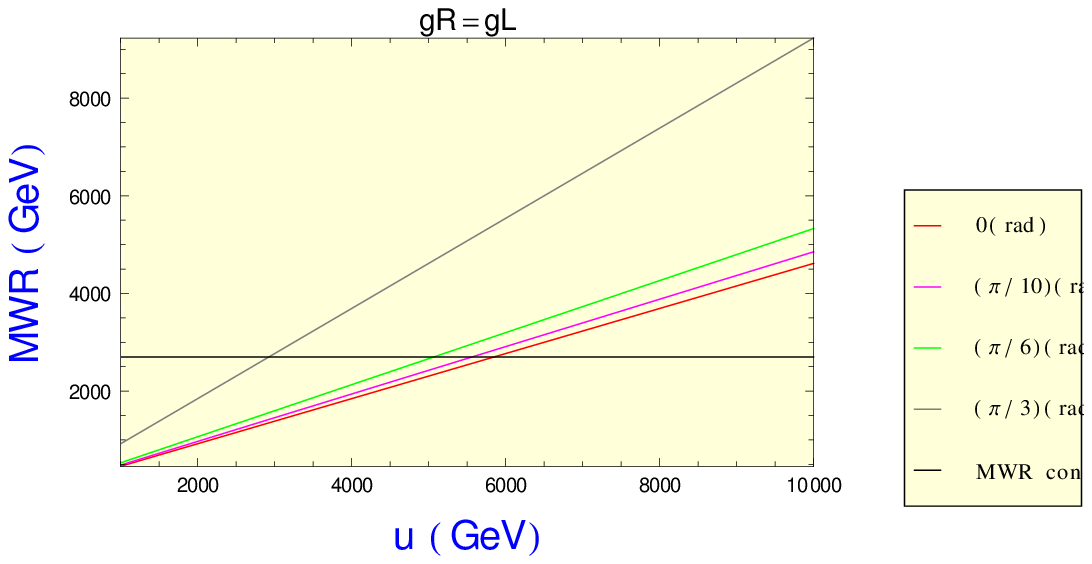,width=0.7
\textwidth,angle=0}}       
\end{center}
\caption{The masses of $W^{\pm}_{R}$ to several values of the 
$\alpha$ parameter as in Fig.(\ref{fig2}).}
\label{fig3}
\end{figure}
\item[3] $g_{R}<g_{L}$
\end{itemize}
\begin{figure}[ht]
\begin{center}
\vglue -0.009cm
\mbox{\epsfig{file=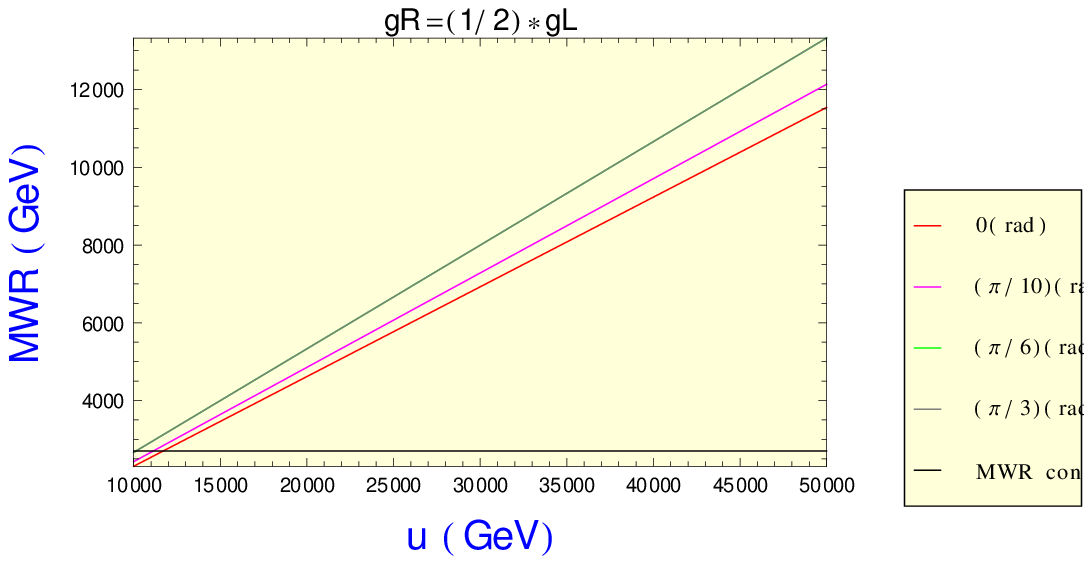,width=0.7\textwidth,angle=0}}       
\end{center}
\caption{The masses of $W^{\pm}_{R}$ to several values of the 
$\alpha$ parameter 
as in Fig.(\ref{fig2}).}
\label{fig4}
\end{figure}

Now the plot of the masses of 
$W^{\pm}_{R}$ is shown at 
Figs.(\ref{fig2},\ref{fig3},\ref{fig4}) and 
they are in agreement with 
Eq.(\ref{mwlbigmwr}), see also 
Fig.(\ref{fig1}), for $v_{R}> 2000$ GeV.

\subsection{Neutral Gauge Bosons}
The mass matrix to the neutral gauge boson at the base $(V^{3L}_{m},V^{3R}_{m},
V^{BL}_{m})$ is given by
\begin{equation}
{\cal M}^{2}_{neu}=
\left( 
\begin{array}{ccc} 
g^{2}_{L}(v_{L}^{2}+v_{L}^{\prime \,\ 2}) & 0 & -g_{BL}g_{L}(v_{L}^{2}+v_{L}^{\prime \,\ 2}) \\ 
0 & g^{2}_{R}(v_{R}^{2}+v_{R}^{\prime \,\ 2}) & -g_{BL}g_{R}(v_{R}^{2}+v_{R}^{\prime \,\ 2}) \\
-g_{BL}g_{L}(v_{L}^{2}+v_{L}^{\prime \,\ 2}) & -g_{BL}g_{R}(v_{R}^{2}+v_{R}^{\prime \,\ 2}) & g^{2}_{BL}(v_{L}^{2}+v_{L}^{\prime \,\ 2}+v_{R}^{2}+v_{R}^{\prime \,\ 2})
\end{array} 
\right),
\label{neutralgagebosonmixing}
\end{equation}
we get analytical the following result
\begin{eqnarray}
{\mbox Det}{\cal M}^{2}_{neu}&=&0, 
\nonumber \\
{\mbox Tr}{\cal M}^{2}_{neu}&=& 
g^{2}_{L}\left( 
v_{L}^{2}+v_{L}^{\prime \,\ 2}
\right)+
g^{2}_{R}\left(
v_{R}^{2}+v_{R}^{\prime \,\ 2}
\right)+
g^{2}_{BL}\left(
v_{L}^{2}+v_{L}^{\prime \,\ 2}+v_{R}^{2}+v_{R}^{\prime \,\ 2}
\right), \nonumber \\
\label{neutralgaugebosonsmasses}
\end{eqnarray}
this result is in agreement with the results 
presented at \cite{Davidson:1987mh}. 

We have one 
non massive gauge boson, the photon, 
and two massive gauge bosons and they are 
$Z^{0}_{m},Z^{\prime 0}_{m}$. The photon is defined as
\begin{equation}
A^{0}_{m}= \sin \theta V^{3L}_{m}+ \cos \theta \left( 
\sin \xi V^{3R}_{m}+ \cos 
\xi V^{BL}_{m} 
\right),
\end{equation}
where we have defined
\begin{eqnarray}
\sin \theta &=& g_{BL}g_{R},
\,\
\cos \theta = 
\sqrt{1-g^{2}_{BL}g^{2}_{R}},
\nonumber \\
\sin \xi &=& \frac{g_{BL}g_{L}}{\sqrt{1-g^{2}_{BL}g^{2}_{R}}},
\,\
\cos \xi = \frac{g_{L}g_{R}}{\sqrt{1-g^{2}_{BL}g^{2}_{R}}}.
\end{eqnarray}

The gauge coupling must be 
related by
\begin{eqnarray}
\frac{1}{e^{2}}=
\frac{1}{g^{2}_{L}}+
\frac{1}{g^{2}_{R}}+
\frac{1}{g^{2}_{BL}},
\end{eqnarray}
therefore, we can conclude
\begin{equation}
g_{BL}=\frac{eg_{L}g_{R}}{\sqrt{g^{2}_{L}g^{2}_{R}-
e^{2}g^{2}_{R}-
e^{2}g^{2}_{L}}}.
\end{equation}
Using Eq.(\ref{gSM}) together with the equation above we get
\begin{eqnarray}
g_{L}&=&g_{R}=0.652775, \nonumber \\
g_{BL}&=&0.121505.
\label{gLgRgBL}
\end{eqnarray} 

\begin{figure}[ht]
\begin{center}
\vglue -0.009cm
\mbox{\epsfig{file=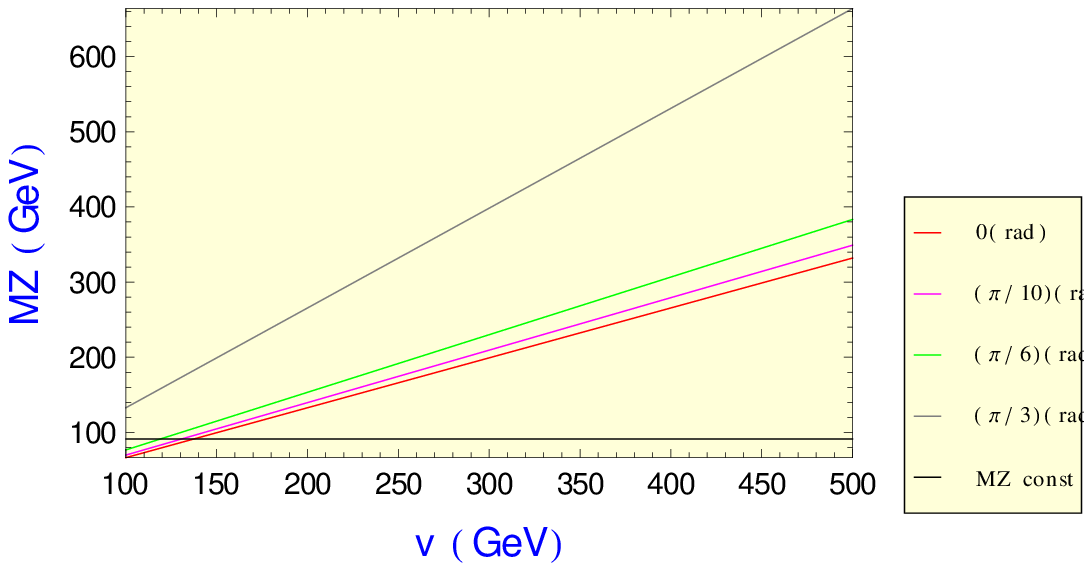,width=0.7
\textwidth,angle=0}}       
\end{center}
\caption{The masses of $Z^{0}$ to several values of the 
$\beta$ parameter 
in therms of the vev of $\phi_{L}$, the black line 
means the experimental constraints of $M_{Z^{0}}$ 
defined at Eq.(\ref{Zmassconstraints}).}
\label{figMZ}
\end{figure}

\begin{figure}[ht]
\begin{center}
\vglue -0.009cm
\mbox{\epsfig{file=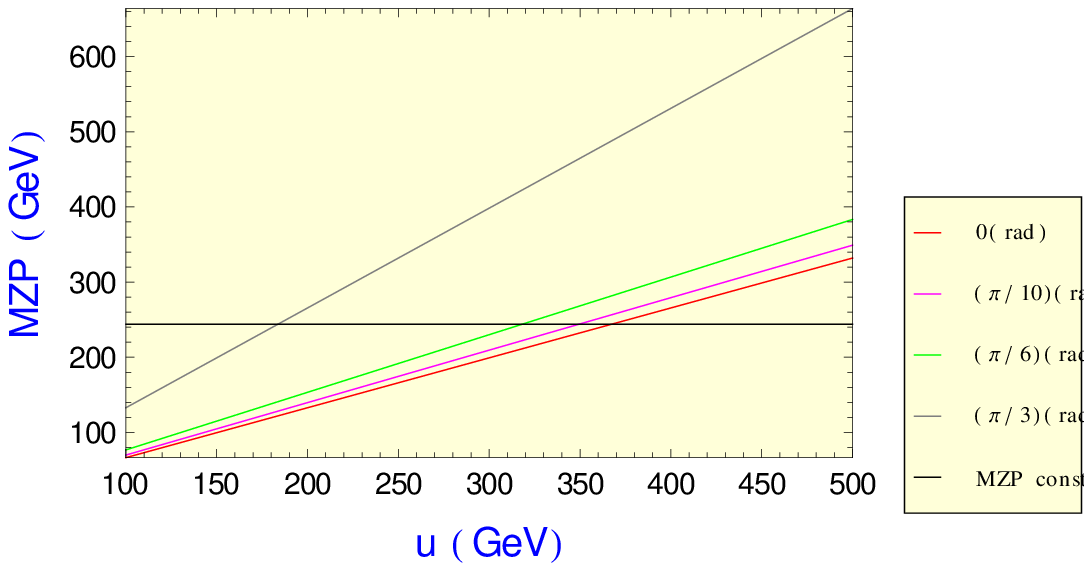,width=0.7
\textwidth,angle=0}}       
\end{center}
\caption{The masses of $Z^{\prime 0}$ to several values of the 
$\alpha$ parameter 
in therms of the vev of $\phi_{R}$, the black line 
means the experimental constraints of $M_{Z^{\prime 0}}$ 
defined at Eq.(\ref{WRmassconstraints}).}
\label{figMZP}
\end{figure}

The analytical results are so complicated, due it we will not reproduce it here. However, we have done some plots about the masses of both massive gauge bosons and our results are shown at Figs.(\ref{figMZ},\ref{figMZP}). There are an experimental 
constraints for the masses of 
$Z^{\prime 0}$ and it is \cite{Zyla:2020zbs} 
\begin{equation}
M_{Z^{\prime 0}}>1 \,\ 
{\mbox TeV}.
\label{ZPmassconstraints}
\end{equation}
for the case of 
$g_{L}=g_{R}$, while the experimental value of $Z^{0}$-mass is given by:
\begin{equation}
M_{Z^{0}}=91.1876 \pm 0.0021 \,\  
{\mbox GeV}.
\label{Zmassconstraints}
\end{equation}

\section{Scalar Potential}

In the non-SUSY model LRUSM, the scalar 
potential is given by \cite{Davidson:1987mh}:
\begin{eqnarray}
V&=&n^{2}_{L} \vert \phi_{L} \vert^{2} +
n^{2}_{R} \vert \phi_{R} \vert^{2} +
\lambda_{LL}\left( \vert \phi_{L} \vert^{2} 
\right)^{2} +
\lambda_{RR}\left( \vert \phi_{R} \vert^{2} 
\right)^{2} +
\lambda_{LR}\left( \vert \phi^{\dagger}_{L} 
\phi_{R} \vert^{2} \right)^{2}. \nonumber \\
\end{eqnarray}
One year later they also included one singlet scalar field\footnote{They call this singlet as $\sigma$ and in our notatio it is $S$.}, see 
Eq.(\ref{newsingletscalar}), and their main resultas are \cite{Davidson:1987tr}:
\begin{itemize}
\item The Dine-Fischler-Srednicki role;
\item The Chang-Mohapatra-Parida role;
\item The Universal Seesaw Mechanism.
\end{itemize}
for the last result, see 
Eq.(\ref{vevofthemodel})

The scalar potential is written as
\begin{equation}
V_{MSUSM}=V_{D}+V_{F}+V_{\mbox{soft}}
\label{ep1}
\end{equation}
where, see Eqs.(\ref{dtoscalarpotential},\ref{ftoscalarpotential},\ref{smtsoft}), we can rewrite
\begin{eqnarray}
V_{D}&=&-{\cal L}_{D}=\frac{1}{2}\left(
D^{i}_{L}D^{i}_{L}+
D^{i}_{R}D^{i}_{R}+
D^{BL}D^{BL}\right)\nonumber \\ &=&\frac{g^{2}_{L}}{8} \left[
\left(
\vert \phi_{L}\vert^{2}-
\vert \phi^{\prime}_{L}\vert^{2}
\right)^{2}+4
\vert 
\phi^{\dagger \,\ \prime}_{L} 
\phi_{L}
\vert^{2}
\right] +
\frac{g^{2}_{R}}{8} \left[
\left(
\vert \phi_{R}\vert^{2}-
\vert \phi^{\prime}_{R}\vert^{2}
\right)^{2}+4
\vert 
\phi^{\dagger \,\ \prime}_{R} 
\phi_{R}
\vert^{2}
\right] 
\nonumber \\
&+& 
\frac{g^{2}_{BL}}{8}\left\{
\left[ \left( 
\vert \phi_{L}\vert^{2}-
\vert \phi^{\prime}_{L}\vert^{2}
\right)+
\left( 
\vert \phi_{R}\vert^{2}-
\vert \phi^{\prime}_{R}\vert^{2}
\right) \right]^{2}+ 
\left( \vert S \vert^{2}\right)^{2}-2
\left( 
\vert \phi_{L}\vert^{2}-
\vert \phi^{\prime}_{L}\vert^{2}
\right) \vert S \vert^{2}
\right. \nonumber \\
&-& \left. 2 
\left( 
\vert \phi_{R}\vert^{2}-
\vert \phi^{\prime}_{R}\vert^{2}
\right) \vert S \vert^{2}
\right\}, 
\nonumber \\
V_{F}&=&-{\cal L}_{F}=\sum_{m}\bar{F}_{m}F_{m}\nonumber \\ &=&
\vert \mu_{L} \vert^{2} 
\vert \phi^{\prime}_{L} \vert^{2}+
\vert \mu_{R} \vert^{2}
\vert \phi^{\prime}_{R} \vert^{2}+
\vert \mu_{L} \vert^{2} 
\vert \phi_{L} \vert^{2}+
\vert \mu_{R} \vert^{2} 
\vert \phi_{R} \vert^{2}+
\vert \mu_{S} \vert^{2} 
\vert S \vert^{2}, 
\nonumber \\
V_{soft}&=&-{\cal L}^{\mbox{scalar}}_{soft}
\nonumber \\ 
&=&M^{2}_{\phi_{L}}
\vert \phi_{L} \vert^{2}+
M^{2}_{\phi_{R}}
\vert \phi_{R} \vert^{2}+
M^{2}_{\phi^{\prime}_{L}}
\vert \phi^{\prime}_{L}
\vert^{2}+
M^{2}_{\phi^{\prime}_{R}}
\vert \phi^{\prime}_{R}
\vert^{2}+
M^{2}_{S}\vert S \vert^{2}  
- \left[
\beta_{\phi_{L}} \left( 
\phi_{L} \phi^{\prime}_{L} 
\right) \right. \nonumber \\
&+& \left. 
\beta_{\phi_{R}} \left( 
\phi_{R} \phi^{\prime}_{R} 
\right) +
\beta_{S} \left( 
S \right)^{2}+ H.c. \right] .
\label{ess}
\end{eqnarray}

Therefore our scalar potential is given by:
\begin{eqnarray}
V_{MSUSM}&=&
m^{2}_{\phi_{L}}
\vert \phi_{L} \vert^{2}+
m^{2}_{\phi_{R}}
\vert \phi_{R} \vert^{2}+
m^{2}_{\phi^{\prime}_{L}}
\vert \phi^{\prime}_{L} \vert^{2}+
m^{2}_{\phi^{\prime}_{R}}
\vert \phi^{\prime}_{R} \vert^{2}+
m^{2}_{S}
\vert S \vert^{2}  
- \left[
\beta_{\phi_{L}} \left( 
\phi_{L} \phi^{\prime}_{L} 
\right) \right. \nonumber \\
&+& \left. 
\beta_{\phi_{R}} \left( 
\phi_{R} \phi^{\prime}_{R} 
\right) +
\beta_{S} \left( 
S \right)^{2}+ H.c. \right]
+ \frac{1}{8}
\left( g^{2}_{L}+g^{2}_{BL} 
\right) \left(
\vert \phi_{L}\vert^{2}-
\vert \phi^{\prime}_{L}\vert^{2}
\right)^{2}
\nonumber \\
&+& \frac{1}{8}
\left( g^{2}_{R}+g^{2}_{BL} 
\right) \left(
\vert \phi_{R}\vert^{2}-
\vert \phi^{\prime}_{R}\vert^{2}
\right)^{2}
+ \frac{g^{2}_{L}}{2}
\vert 
\phi^{\dagger \,\ \prime}_{L} 
\phi_{L} \vert^{2}
+ \frac{g^{2}_{R}}{2}
\vert 
\phi^{\dagger \,\ \prime}_{R} 
\phi_{R} \vert^{2}
\nonumber \\
&+&
\frac{g^{2}_{BL}}{8}
\left[
2\left( 
\vert \phi_{L}\vert^{2}-
\vert \phi^{\prime}_{L}\vert^{2}
\right) \left( 
\vert \phi_{R}\vert^{2}-
\vert \phi^{\prime}_{R}\vert^{2}
\right)+
\left( \vert S \vert^{2}\right)^{2}-2
\left( 
\vert \phi_{L}\vert^{2}-
\vert \phi^{\prime}_{L}\vert^{2}
\right) \vert S \vert^{2}
\right. \nonumber \\
&-& \left. 2 
\left( 
\vert \phi_{R}\vert^{2}-
\vert \phi^{\prime}_{R}\vert^{2}
\right) \vert S \vert^{2}
\right] .
\label{ep1}
\end{eqnarray}
where
\begin{eqnarray}
m^{2}_{\phi_{L}}&=& 
M^{2}_{\phi_{L}}+ 
\vert \mu_{L} \vert^{2}, \,\
m^{2}_{\phi_{R}}=
M^{2}_{\phi_{R}}+
\vert \mu_{R} \vert^{2}, \nonumber \\
m^{2}_{\phi^{\prime}_{L}}&=&
M^{2}_{\phi^{\prime}_{L}}+
\vert \mu_{L} \vert^{2}, \,\
m^{2}_{\phi^{\prime}_{R}}=
M^{2}_{\phi^{\prime}_{R}}+
\vert \mu_{R} \vert^{2}, \,\
m^{2}_{S}=
M^{2}_{S}+
\vert \mu_{S} \vert^{2}.
\end{eqnarray}

All the five neutral scalar components 
$\phi^{0}_{L},\phi^{0}_{R},\phi^{\prime 0}_{L},\phi^{\prime 0}_{R},S$
gain non-zero vacuum expectation values. Making a shift in 
the neutral scalars as
\begin{eqnarray}
< \phi_{L} > &=& 
\frac{1}{\sqrt{2}} 
      \left( \begin{array}{c}
v_{L}+H_{\phi_{L}}+ \imath
F_{\phi_{L}}
\\ 
                  0  \end{array} \right) 
\,\ , \,\
< \phi^{\prime}_{L} > = 
\frac{1}{\sqrt{2}} 
      \left( \begin{array}{c}
      0 \\
v^{\prime}_{L}
+H_{\phi^{\prime}_{L}}+ \imath
F_{\phi^{\prime}_{L}}

\end{array} \right) 
\,\ ,  \nonumber \\
< \phi_{R} > &=& 
\frac{1}{\sqrt{2}} 
      \left( \begin{array}{c}
v_{R}+H_{\phi_{R}}+ \imath
F_{\phi_{R}}
\\ 
                  0  \end{array} \right) 
\,\ , \,\
< \phi^{\prime}_{R} > = 
\frac{1}{\sqrt{2}} 
      \left( \begin{array}{c}
      0 \\
v^{\prime}_{R}
+H_{\phi^{\prime}_{R}}+ \imath
F_{\phi^{\prime}_{R}}
\end{array} \right) 
\,\ ,  \nonumber \\
< S > &=& \frac{1}{\sqrt{2}} 
\left( x+H_{S}+ \imath 
F_{S}  
\right).
\label{develop}
\end{eqnarray}

\label{sec:sp}

In the non-SUSY model LRUSM, the scalar 
potential is given by \cite{Davidson:1987mh}:
\begin{eqnarray}
V&=&n^{2}_{L} \vert \phi_{L} \vert^{2} +
n^{2}_{R} \vert \phi_{R} \vert^{2} +
\lambda_{LL}\left( \vert \phi_{L} \vert^{2} 
\right)^{2} +
\lambda_{RR}\left( \vert \phi_{R} \vert^{2} 
\right)^{2} +
\lambda_{LR}\left( \vert \phi^{\dagger}_{L} 
\phi_{R} \vert^{2} \right)^{2}. \nonumber \\
\end{eqnarray}
One year later they also included one singlet scalar field\footnote{They call this singlet as $\sigma$ and in our notatio it is $S$.}, see 
Eq.(\ref{newsingletscalar}), and their main resultas are \cite{Davidson:1987tr}:
\begin{itemize}
\item The Dine-Fischler-Srednicki role;
\item The Chang-Mohapatra-Parida role;
\item The Universal Seesaw Mechanism.
\end{itemize}

The scalar potential is written as
\begin{equation}
V_{MSUSM}=V_{D}+V_{F}+V_{\mbox{soft}}
\label{ep1}
\end{equation}
where, see Eqs.(\ref{dtoscalarpotential},\ref{ftoscalarpotential},\ref{smtsoft}), we can rewrite
\begin{eqnarray}
V_{D}&=&-{\cal L}_{D}=\frac{1}{2}\left(
D^{i}_{L}D^{i}_{L}+
D^{i}_{R}D^{i}_{R}+
D^{BL}D^{BL}\right)\nonumber \\ &=&\frac{g^{2}_{L}}{8} \left[
\left(
\vert \phi_{L}\vert^{2}-
\vert \phi^{\prime}_{L}\vert^{2}
\right)^{2}+4
\vert 
\phi^{\dagger \,\ \prime}_{L} 
\phi_{L}
\vert^{2}
\right] +
\frac{g^{2}_{R}}{8} \left[
\left(
\vert \phi_{R}\vert^{2}-
\vert \phi^{\prime}_{R}\vert^{2}
\right)^{2}+4
\vert 
\phi^{\dagger \,\ \prime}_{R} 
\phi_{R}
\vert^{2}
\right] 
\nonumber \\
&+& 
\frac{g^{2}_{BL}}{8}\left\{
\left[ \left( 
\vert \phi_{L}\vert^{2}-
\vert \phi^{\prime}_{L}\vert^{2}
\right)+
\left( 
\vert \phi_{R}\vert^{2}-
\vert \phi^{\prime}_{R}\vert^{2}
\right) \right]^{2}+ 
\left( \vert S \vert^{2}\right)^{2}-2
\left( 
\vert \phi_{L}\vert^{2}-
\vert \phi^{\prime}_{L}\vert^{2}
\right) \vert S \vert^{2}
\right. \nonumber \\
&-& \left. 2 
\left( 
\vert \phi_{R}\vert^{2}-
\vert \phi^{\prime}_{R}\vert^{2}
\right) \vert S \vert^{2}
\right\}, 
\nonumber \\
V_{F}&=&-{\cal L}_{F}=\sum_{m}\bar{F}_{m}F_{m}\nonumber \\ &=&
\vert \mu_{L} \vert^{2} 
\vert \phi^{\prime}_{L} \vert^{2}+
\vert \mu_{R} \vert^{2}
\vert \phi^{\prime}_{R} \vert^{2}+
\vert \mu_{L} \vert^{2} 
\vert \phi_{L} \vert^{2}+
\vert \mu_{R} \vert^{2} 
\vert \phi_{R} \vert^{2}+
\vert \mu_{S} \vert^{2} 
\vert S \vert^{2}, 
\nonumber \\
V_{soft}&=&-{\cal L}^{\mbox{scalar}}_{soft}
\nonumber \\ 
&=&M^{2}_{\phi_{L}}
\vert \phi_{L} \vert^{2}+
M^{2}_{\phi_{R}}
\vert \phi_{R} \vert^{2}+
M^{2}_{\phi^{\prime}_{L}}
\vert \phi^{\prime}_{L}
\vert^{2}+
M^{2}_{\phi^{\prime}_{R}}
\vert \phi^{\prime}_{R}
\vert^{2}+
M^{2}_{S}\vert S \vert^{2}  
- \left[
\beta_{\phi_{L}} \left( 
\phi_{L} \phi^{\prime}_{L} 
\right) \right. \nonumber \\
&+& \left. 
\beta_{\phi_{R}} \left( 
\phi_{R} \phi^{\prime}_{R} 
\right) +
\beta_{S} \left( 
S \right)^{2}+ H.c. \right] .
\label{ess}
\end{eqnarray}

Therefore our scalar potential is given by:
\begin{eqnarray}
V_{MSUSM}&=&
m^{2}_{\phi_{L}}
\vert \phi_{L} \vert^{2}+
m^{2}_{\phi_{R}}
\vert \phi_{R} \vert^{2}+
m^{2}_{\phi^{\prime}_{L}}
\vert \phi^{\prime}_{L} \vert^{2}+
m^{2}_{\phi^{\prime}_{R}}
\vert \phi^{\prime}_{R} \vert^{2}+
m^{2}_{S}
\vert S \vert^{2}  
- \left[
\beta_{\phi_{L}} \left( 
\phi_{L} \phi^{\prime}_{L} 
\right) \right. \nonumber \\
&+& \left. 
\beta_{\phi_{R}} \left( 
\phi_{R} \phi^{\prime}_{R} 
\right) +
\beta_{S} \left( 
S \right)^{2}+ H.c. \right]
+ \frac{1}{8}
\left( g^{2}_{L}+g^{2}_{BL} 
\right) \left(
\vert \phi_{L}\vert^{2}-
\vert \phi^{\prime}_{L}\vert^{2}
\right)^{2}
\nonumber \\
&+& \frac{1}{8}
\left( g^{2}_{R}+g^{2}_{BL} 
\right) \left(
\vert \phi_{R}\vert^{2}-
\vert \phi^{\prime}_{R}\vert^{2}
\right)^{2}
+ \frac{g^{2}_{L}}{2}
\vert 
\phi^{\dagger \,\ \prime}_{L} 
\phi_{L} \vert^{2}
+ \frac{g^{2}_{R}}{2}
\vert 
\phi^{\dagger \,\ \prime}_{R} 
\phi_{R} \vert^{2}
\nonumber \\
&+&
\frac{g^{2}_{BL}}{8}
\left[
2\left( 
\vert \phi_{L}\vert^{2}-
\vert \phi^{\prime}_{L}\vert^{2}
\right) \left( 
\vert \phi_{R}\vert^{2}-
\vert \phi^{\prime}_{R}\vert^{2}
\right)+
\left( \vert S \vert^{2}\right)^{2}-2
\left( 
\vert \phi_{L}\vert^{2}-
\vert \phi^{\prime}_{L}\vert^{2}
\right) \vert S \vert^{2}
\right. \nonumber \\
&-& \left. 2 
\left( 
\vert \phi_{R}\vert^{2}-
\vert \phi^{\prime}_{R}\vert^{2}
\right) \vert S \vert^{2}
\right] .
\label{ep1}
\end{eqnarray}
where
\begin{eqnarray}
m^{2}_{\phi_{L}}&=& 
M^{2}_{\phi_{L}}+ 
\vert \mu_{L} \vert^{2}, \,\
m^{2}_{\phi_{R}}=
M^{2}_{\phi_{R}}+
\vert \mu_{R} \vert^{2}, \nonumber \\
m^{2}_{\phi^{\prime}_{L}}&=&
M^{2}_{\phi^{\prime}_{L}}+
\vert \mu_{L} \vert^{2}, \,\
m^{2}_{\phi^{\prime}_{R}}=
M^{2}_{\phi^{\prime}_{R}}+
\vert \mu_{R} \vert^{2}, \,\
m^{2}_{S}=
M^{2}_{S}+
\vert \mu_{S} \vert^{2}.
\end{eqnarray}

All the five neutral scalar components 
$\phi^{0}_{L},\phi^{0}_{R},\phi^{\prime 0}_{L},\phi^{\prime 0}_{R},S$
gain non-zero vacuum expectation values. Making a shift in 
the neutral scalars as
\begin{eqnarray}
< \phi_{L} > &=& 
\frac{1}{\sqrt{2}} 
      \left( \begin{array}{c}
v_{L}+H_{\phi_{L}}+ \imath
F_{\phi_{L}}
\\ 
                  0  \end{array} \right) 
\,\ , \,\
< \phi^{\prime}_{L} > = 
\frac{1}{\sqrt{2}} 
      \left( \begin{array}{c}
      0 \\
v^{\prime}_{L}
+H_{\phi^{\prime}_{L}}+ \imath
F_{\phi^{\prime}_{L}}

\end{array} \right) 
\,\ ,  \nonumber \\
< \phi_{R} > &=& 
\frac{1}{\sqrt{2}} 
      \left( \begin{array}{c}
v_{R}+H_{\phi_{R}}+ \imath
F_{\phi_{R}}
\\ 
                  0  \end{array} \right) 
\,\ , \,\
< \phi^{\prime}_{R} > = 
\frac{1}{\sqrt{2}} 
      \left( \begin{array}{c}
      0 \\
v^{\prime}_{R}
+H_{\phi^{\prime}_{R}}+ \imath
F_{\phi^{\prime}_{R}}
\end{array} \right) 
\,\ ,  \nonumber \\
< S > &=& \frac{1}{\sqrt{2}} 
\left( x+H_{S}+ \imath 
F_{S}  
\right).
\label{develop}
\end{eqnarray}

\section{Constraint Equations}
\label{sec:constraint}

Here in this section we give the constraint equations, due to the requirement 
the potential to reach a minimum at the chosen VEV's. We get this equation 
requiring that in the shifted potential the linear terms in fields must 
be absent
\begin{eqnarray}
8v_{L}M^{2}_{\phi_{L}}&=& 
8 \beta_{\phi_{L}}
v^{\prime}_{L}+v_{L} \left[
(g^{2}_{L}+g^{2}_{BL})
(v^{\prime 2}_{L}-v^{2}_{L})
+g^{2}_{BL}
(v^{\prime 2}_{R}-v^{2}_{R}) 
+8(x^{2}-\mu^{2}_{L})
\right], 
\nonumber \\
8v_{R}M^{2}_{\phi_{R}}&=&- 
8 \beta_{\phi_{R}}
v^{\prime}_{R}+v_{R} \left[
(g^{2}_{R}+g^{2}_{BL})
(v^{\prime 2}_{R}-v^{2}_{R})
+g^{2}_{BL}
(v^{\prime 2}_{L}-v^{2}_{L}) 
+8(x^{2}-\mu^{2}_{R})
\right],
\nonumber \\
8v^{\prime}_{L}
M^{2}_{\phi^{\prime}_{L}}
&=&- 8 \beta_{\phi_{L}}
v_{L}-v^{\prime}_{L} \left[
g^{2}_{BL}
(v^{\prime 2}_{R}-v^{2}_{R})+ 
(g^{2}_{L}+g^{2}_{BL})
(v^{\prime 2}_{L}-v^{2}_{L}) 
-8(x^{2}+\mu^{2}_{L})
\right],
\nonumber \\
8v^{\prime}_{R}
M^{2}_{\phi^{\prime}_{R}}
&=& 8 \beta_{\phi_{R}}
v_{R}+v^{\prime}_{R} \left[
g^{2}_{BL}
(v^{2}_{L}-v^{\prime 2}_{L})+ 
(g^{2}_{R}+g^{2}_{BL})
(v^{2}_{R}-v^{\prime 2}_{R}) 
-8(x^{2}+\mu^{2}_{R}) 
\right],
\nonumber \\
M^{2}_{S}
&=& 2 \beta_{S}+v^{2}_{L}
+v^{2}_{R}-v^{\prime 2}_{L}
-v^{\prime 2}_{R}-x^{2}
-\mu^{2}_{S}.
\label{constraintequationsinmodel1}
\end{eqnarray}

The mass matrices, thus, can be calculated, using
\begin{equation}
M_{ij}^2=\frac{\partial ^2V_{MSUSY331}}{\partial \phi _i\partial \phi _j}
\label{calculomassamodelo1}
\end{equation}
evaluated at the chosen minimum, where $\phi_i$ are the scalars of our 
model described above.

For the sake of simplicity, here we assume that 
vacuum expectation values (VEVs) are real. This means that
the CP violation through the scalar exchange is not considered
in this work. In literature, a real part $H$ is called  CP-even scalar or 
{\it scalar}, and an imaginary one  $F$  -  CP-odd scalar or 
{\it pseudoscalar} field. In this paper we call them  scalar and pseudoscalar, 
respectively.

\section{Mass Spectrum general case}
\label{sec:analyses}

We can write, from 
Eqs.(\ref{defbetaalpha},
\ref{hlmassexpression},
\ref{eigenvaluescpodd},
\ref{eigenvaluescharged}), 
the following expression for our light $CP$-even, $CP$-odd and Charged scalars:
\begin{eqnarray}
M_{H^{0}_{1}}&=&
\sqrt{\left( 
\frac{g^{2}_{L}+g^{2}_{BL}}{4}
\right) v^{2}_{L} \sec \beta +
\left( 
\frac{g^{2}_{R}+g^{2}_{BL}}{4}
\right) v^{2}_{R} \sec \alpha}, 
\nonumber \\
m_{A^{0}_{1}}&=& \beta_{\phi_{L}}
\sqrt{ 
\frac{2}{ \sin \left( 2 \beta \right)}},
\,\ 
m_{A^{0}_{2}}= \beta_{\phi_{R}}
\sqrt{ 
\frac{2}{ \sin \left( 2 \alpha \right)}}, 
\nonumber \\
m_{H^{\pm}_{1}}&=&
\sqrt{
\left[ 
\frac{\beta^{2}_{\phi_{L}}}{\tan \beta} 
+ 
\frac{g^{2}_{L}}{4} v^{2}_{L} 
\right] \sec \beta
}. 
\label{allscalarslightmassexpression}
\end{eqnarray}
We have discussed that the masses of those 
light scalars are very similar to ones at MSSM \cite{Rodriguez:2019mwf}. The 
$\beta$-parameter has to satisfy the 
following constraints
\begin{eqnarray}
\beta &\neq& 0 \,\ \mbox{rad}, \,\ \mbox{and} \,\
\beta \neq \frac{\pi}{2} \,\ \mbox{rad}, 
\nonumber \\
0 &<& \beta < \frac{\pi}{2} \,\ \mbox{rad},
\label{constraintsinbetaparameter}
\end{eqnarray}
and $\alpha$-parameter also has to satisfy 
those condition above presented. 
However on this model the following 
relations hold in MSSM \cite{Rodriguez:2019mwf}
\begin{eqnarray}
m^{2}_{H^{\pm}}&=&m^{2}_{A^{0}}+
M^{2}_{W^{\pm}}, 
\nonumber \\
m^{2}_{h^{0}}+m^{2}_{H^{0}}&=&
m^{2}_{A^{0}}+M^{2}_{Z^{0}}.
\end{eqnarray}
are not hold in our model.

Using Eqs.(\ref{gSM},\ref{gLgRgBL}) 
at Eqs.(\ref{eigenvaluescpodd},\ref{eigenvaluescharged}) and 
we use $\beta_{\phi_{L}}=2000$ GeV; 
$\beta_{\phi_{R}}=3000$ GeV 
and $\beta_{S}=2500$ GeV; we get 
the following masses values for the masses of our scalars:
\begin{eqnarray}
m_{A^{0}_{1}}&=&3039.34 \,\ \mbox{GeV}, \,\
m_{A^{0}_{2}}=4559.01 \,\ \mbox{GeV}, \,\
m_{A^{0}_{3}}=5000 \,\ \mbox{GeV}
, \nonumber \\
m_{H^{\pm}_{1}}&=&3039.68 \,\ \mbox{GeV}, \,\
m_{H^{\pm}_{2}}=4561.93 \,\ \mbox{GeV},
\end{eqnarray}
those values are in agreement with 
the experimental limits for new scalars 
defined as 
\begin{eqnarray}
m_{A^{0}}&>&863 \,\ \mbox{GeV}, 
\,\
m_{H^{+}}>181 \,\ \mbox{GeV}, 
\nonumber \\
\label{expresultas}
\end{eqnarray}
at $\tan \beta =10$ \cite{Zyla:2020zbs}.

We have done at Fig.(\ref{fig01}) plots 
for $\beta = 0, ( \pi /10), ( \pi /6), 
( \pi /3)$ rad but considerating the 
Eq.(\ref{constraintsinbetaparameter}) we 
will not use the first values and due the fact that 
$\tan \left( \frac{\pi}{6} \right) = 
\cot \left( \frac{\pi}{3} \right)$ we will 
also not present the last value for $\beta$ 
and $\alpha$-parameters when we present our plots to the scalars masses. 

\begin{figure}[ht]
\begin{center}
\vglue -0.009cm
\mbox{\epsfig{file=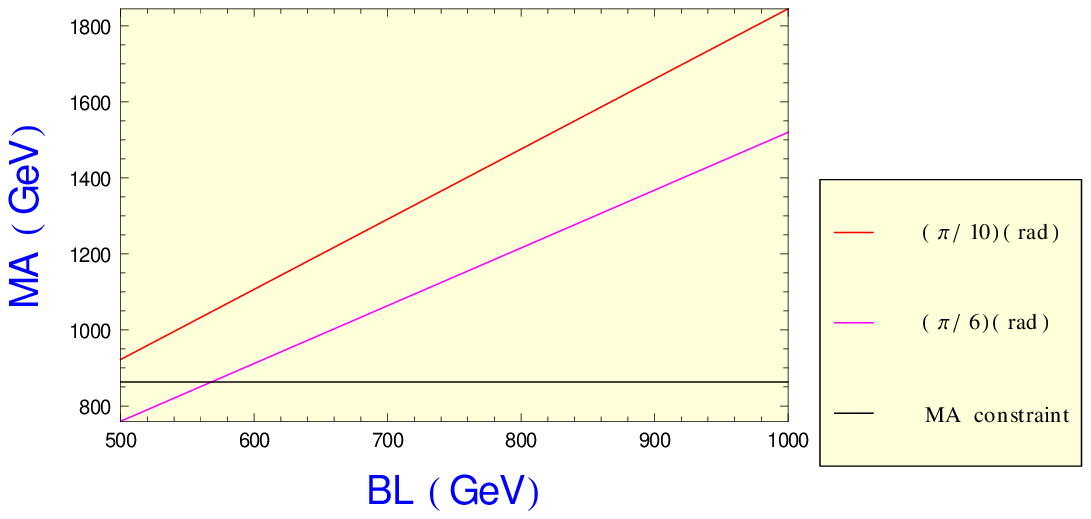,width=0.7
\textwidth,angle=0}}       
\end{center}
\caption{The masses of $A$ to several values of the 
$\beta$ parameter, for $\beta = ( \pi /10), ( \pi /6)$ rad and $v_{L}=120$ GeV; 
in therms of the soft parameter of $L \def \beta_{\phi_{L}}$. The black line 
means the experimental constraints of $M_{A}$ 
defined at Eq.(\ref{expresultas}).}
\label{figMA}
\end{figure}

\begin{figure}[ht]
\begin{center}
\vglue -0.009cm
\mbox{\epsfig{file=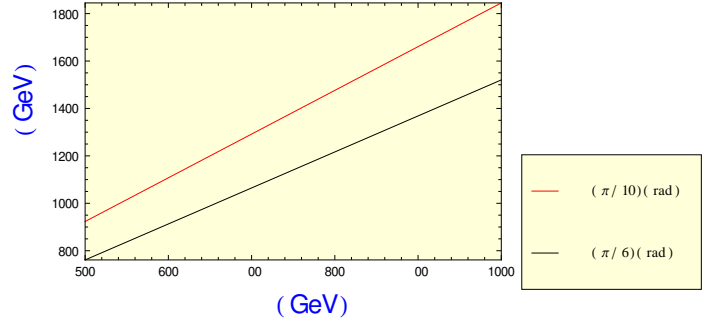,width=0.7
\textwidth,angle=0}}       
\end{center}
\caption{The masses of $H^{\pm}$ to several values of the 
$\beta$ parameter in similar way as done 
at Fig.(\ref{figMA}).}
\label{figMHC}
\end{figure}

We show at Figs.(\ref{figMA},\ref{figMHC}) the masses of pseudo and charged Higgs as funtion 
of $\beta_{\phi_{L}}$. We can conclude when 
we consider $v_{L}=120$ GeV we can get masses four our light  
pseudo-scalar and also charged scalar bigger than the experimental constraints.

\begin{figure}[ht]
\begin{center}
\vglue -0.009cm
\mbox{\epsfig{file=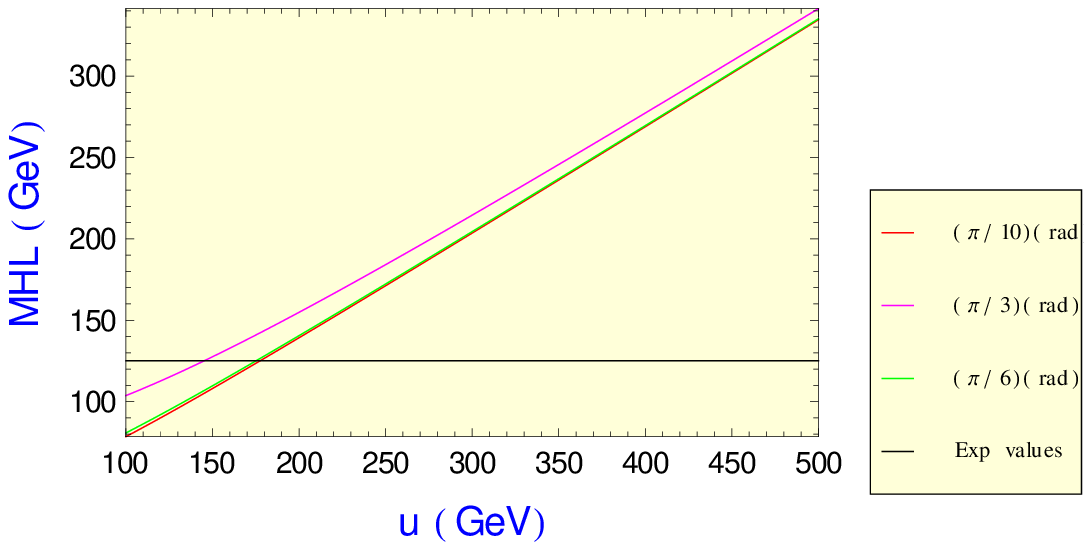,width=0.7
\textwidth,angle=0}}       
\end{center}
\caption{The masses of $H^{0}_{1}$ to several values of the 
$\beta$ parameter and $\alpha = (\pi/3)$ 
in therms of the soft parameter of $v_{R} \equiv u$, the black line 
means the experimental constraints of $M_{H^{0}_{1}}$ 
defined at Eq.(\ref{{hlexpvalue}}).}
\label{figmhlu}
\end{figure}

\begin{figure}[ht]
\begin{center}
\vglue -0.009cm
\mbox{\epsfig{file=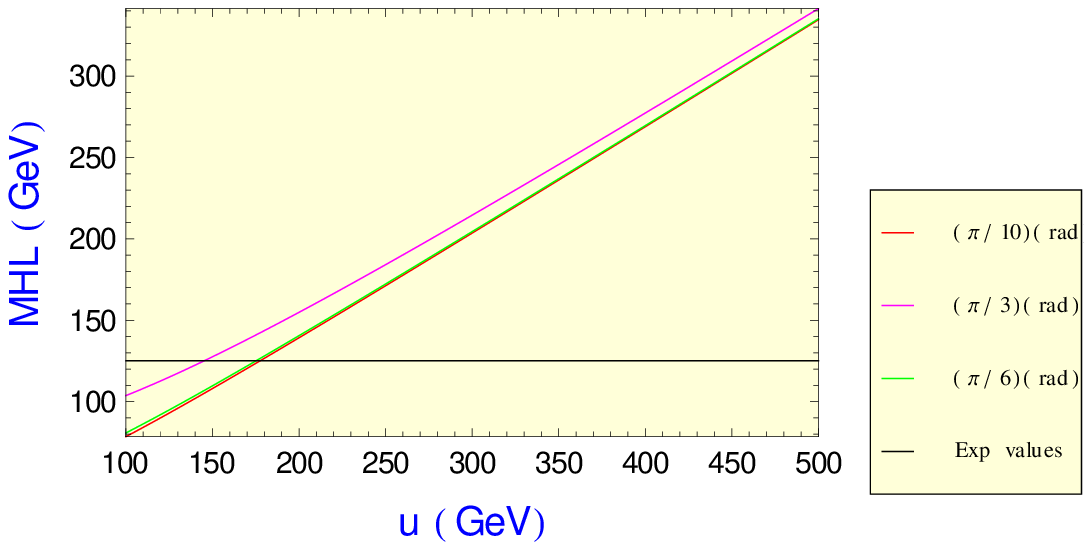,width=0.7
\textwidth,angle=0}}       
\end{center}
\caption{The masses of $H^{0}_{1}$ to several values of the 
$\alpha$ parameter and $\beta = (\pi/3)$ 
in therms of the soft parameter of $v_{L} \equiv v$, the black line 
means the experimental constraints of $M_{H^{0}_{1}}$ 
defined at Eq.(\ref{{hlexpvalue}}).}
\label{figmhlv}
\end{figure}

Our results for the light Higgs bosons are shown at Figs.(\ref{figmhlu},\ref{figmhlv}), and we see that we can reproduce for our light Higgs boson the experimental values of light Higgs Boson is 
\cite{Zyla:2020zbs}:
\begin{equation}
m_{H^{0}_{1}}=125.10 \pm 0.14 \,\ \mbox{GeV}.
\label{hlexpvalue} 
\end{equation}

\section{Conclusions}
\label{sec:conclusion}

In this article we constructed the Supersymmetric 
version of the Universal Seesaw Mechanism. We have presented the model, considerating only one familly fermion coupling with the usual Scalars 
in such way we can explain the mixing data. We have also performed numerical analyses of the gauge boson masses and we also have get the masses for all usual scalar particles. All the mass spectrum are in agreement with the experimental limits, as 
discussed above.


\appendix

\section{Eliminating the Auxiliar Fields}
\label{sec:append1}

To get the scalar potential of our model we have to eliminate the auxiliarly 
fields $F$ and $D$ that appear in our model. We are going to pick up the $F$ 
and $D$- terms, from Eqs.(\ref{l6},\ref{l7},\ref{sp3m1}), we get
\begin{eqnarray}
{\cal L}^{gauge}_{D}&=&
\frac{1}{2} \left( D^{i}_{L}D^{i}_{L}+
D^{i}_{R}D^{i}_{R}+
D^{BL}D^{BL} \right), \nonumber \\
{\cal L}^{scalar}_{F}&=& \vert F_{\phi_{L}} \vert^{2}+
\vert F_{\phi_{R}} \vert^{2}+
\vert F_{\phi^{\prime}_{L}} \vert^{2}+
\vert F_{\phi^{\prime}_{R}} \vert^{2}+
\vert F_{S} \vert^{2}, \nonumber \\
{\cal L}^{scalar}_{D}&=& \frac{g_{L}}{2} \left[
\bar{\phi}_{L}\sigma^{i}
\phi_{L}+
\bar{\phi}^{\prime}_{L}
\sigma^{i}\phi^{\prime}_{L} 
\right]D^{i}_{L} +
\frac{g_{R}}{2} \left[
\bar{\phi}_{R}\sigma^{i}
\phi_{R}+
\bar{\phi}^{\prime}_{R}
\sigma^{i}\phi^{\prime}_{R} 
\right]D^{i}_{R} 
\nonumber \\
&+&
\frac{g_{BL}}{2} \left[
- \vert \phi_{L}\vert^{2}-
\vert \phi_{R}\vert^{2}+
\vert \phi^{\prime}_{L}
\vert^{2}+
\vert \phi^{\prime}_{R}
\vert^{2}+
\vert F_{S}\vert{2} 
\right]D^{BL}, \nonumber \\
{\cal L}^{W2}_{F}&=&
\mu_{L}\left( 
\phi_{L}
F_{\phi^{\prime}_{L}}+
F_{\prime_{L}}
\phi^{\prime}_{L}+
\bar{\phi}_{L}
\bar{F}_{\phi^{\prime}_{L}}+
\bar{F}_{\prime_{L}}
\overline{\phi^{\prime}}_{L} 
\right) \nonumber \\
&+&
\mu_{R}\left( 
\phi_{R}
F_{\phi^{\prime}_{R}}+
F_{\prime_{R}}
\phi^{\prime}_{R}+
\bar{\phi}_{R}
\bar{F}_{\phi^{\prime}_{R}}+
\bar{F}_{\prime_{R}}
\overline{\phi^{\prime}}_{R} 
\right) \nonumber \\
&+&
\mu_{S}\left(
SF_{S}+ \bar{S}\bar{F}_{S} 
\right). 
\nonumber \\
\end{eqnarray}
Our $R$-parity, defined at 
Tab.(\ref{allrpqchargesinMSUSM}), 
eliminate the followings terms in our superpotential
\begin{equation}
\frac{\kappa}{3}\hat{S} 
\hat{S} \hat{S}+
\lambda_{L}\left( 
\hat{\phi}_{L}
\hat{\phi}^{\prime}_{L}
\right) \hat{S} + 
\lambda_{R}\left( 
\hat{\phi}_{R}
\hat{\phi}^{\prime}_{R}
\right) \hat{S}
\end{equation}
those terms would give the following contribute to eliminate the $F$-terms
\begin{eqnarray}
{\cal L}^{W3}_{F}&=& 
\kappa F_{S}S^{2}+
\lambda_{L}\left(
F_{\phi_{L}}
\phi^{\prime}_{L}S+
\phi_{L}F_{\phi^{\prime}_{L}}
S+
\phi_{L}\phi^{\prime}_{L}
F_{S} \right) 
\nonumber \\
&+&
\lambda_{R}\left(
F_{\phi_{R}}
\phi^{\prime}_{R}S+
\phi_{R}F_{\phi^{\prime}_{R}}
S+
\phi_{R}\phi^{\prime}_{R}
F_{S} \right). 
\end{eqnarray}

From the equation described above we can construct
\begin{eqnarray}
{\cal L}_{F}&=&{\cal L}^{scalar}_{F}+{\cal L}^{W2}_{F}
\nonumber \\
&=&
\vert F_{\phi_{L}} \vert^{2}+
\vert F_{\phi_{R}} \vert^{2}+
\vert F_{\phi^{\prime}_{L}} \vert^{2}+
\vert F_{\phi^{\prime}_{R}} \vert^{2}+
\vert F_{S} \vert^{2}+
\mu_{L}\left( 
\phi_{L}
F_{\phi^{\prime}_{L}}+
F_{\prime_{L}}
\phi^{\prime}_{L}+
\bar{\phi}_{L}
\bar{F}_{\phi^{\prime}_{L}}+
\bar{F}_{\prime_{L}}
\overline{\phi^{\prime}}_{L} 
\right) \nonumber \\
&+&
\mu_{R}\left( 
\phi_{R}
F_{\phi^{\prime}_{R}}+
F_{\prime_{R}}
\phi^{\prime}_{R}+
\bar{\phi}_{R}
\bar{F}_{\phi^{\prime}_{R}}+
\bar{F}_{\prime_{R}}
\overline{\phi^{\prime}}_{R} 
\right) \nonumber \\
&+&
\mu_{S}\left(
SF_{S}+ \bar{S}\bar{F}_{S} 
\right)  \,\ , \nonumber \\
{\cal L}_{D}&=&{\cal L}^{gauge}_{D}+{\cal L}^{scalar}_{D} \nonumber \\
&=& 
\frac{1}{2} \left( D^{i}_{L}D^{i}_{L}+
D^{i}_{R}D^{i}_{R}+
D^{BL}D^{BL} \right)
+
\frac{g_{L}}{2} \left[
\bar{\phi}_{L}\sigma^{i}
\phi_{L}+
\bar{\phi}^{\prime}_{L}
\sigma^{i}\phi^{\prime}_{L} 
\right]D^{i}_{L} 
\nonumber \\ &+&
\frac{g_{R}}{2} \left[
\bar{\phi}_{R}\sigma^{i}
\phi_{R}+
\bar{\phi}^{\prime}_{R}
\sigma^{i}\phi^{\prime}_{R} 
\right]D^{i}_{R}+
\frac{g_{BL}}{2} \left[
- \vert \phi_{L}\vert^{2}-
\vert \phi_{R}\vert^{2}+
\vert \phi^{\prime}_{L}
\vert^{2}+
\vert \phi^{\prime}_{R}
\vert^{2}+
\vert S \vert{2} 
\right]D^{BL}. \nonumber \\
\label{auxiliarm1}
\end{eqnarray}

We will now show that these fields can be eliminated through the 
Euler-Lagrange equations
\begin{eqnarray}
\frac{\partial {\cal L}}{\partial \phi}- \partial_{m} 
\frac{\partial {\cal L}}{\partial (\partial_{m} \phi)}=0 \,\ ,  
\label{Euler-Lagrange Equation}
\end{eqnarray}
where 
$\phi = \phi_{L} , \phi_{R} , \phi^{\prime}_{L}, \phi^{\prime}_{R}, S$. 
Formally auxiliary fields are defined 
as fields having no kintetic terms. Thus, this definition immediately yields 
that the Euler-Lagrange equations for auxiliary fields simplify to 
$\frac{\partial {\cal L}}{\partial \phi}=0$.

Applying these simplified equations to various auxiliary $F$-fields yields 
the following relations
\begin{eqnarray}
\bar{F}_{\phi_{L}}&=&- 
\bar{\mu}_{L}\phi^{\prime}_{L}
\,\ ;
\,\ 
F_{\phi_{L}}=- 
\mu_{L}
\overline{\phi^{\prime}}_{L}, 
\nonumber \\
\bar{F}_{\phi_{R}}&=&- 
\bar{\mu}_{R}\phi^{\prime}_{R}
\,\ ;
\,\ 
F_{\phi_{R}}=- 
\mu_{R}
\overline{\phi^{\prime}}_{R}, 
\nonumber \\
\bar{F}_{\phi^{\prime}_{L}}&=&- 
\bar{\mu}_{L}\phi_{L}
\,\ ;
\,\ 
F_{\phi^{\prime}_{L}}=- 
\mu_{L}
\bar{\phi}_{L}, 
\nonumber \\
\bar{F}_{\phi^{\prime}_{R}}&=&- 
\mu_{R}\phi_{R}
\,\ ;
\,\ 
F_{\phi^{\prime}_{R}}=- 
\bar{\mu}_{R}
\overline{\phi}_{R}, 
\nonumber \\
\bar{F}_{S}&=&- \mu_{S}S
\,\ ;
\,\ 
F_{S}=- \bar{\mu}_{S}\bar{S} \,\ ,
\end{eqnarray}
using these equations, we can rewrite Eq.(\ref{auxiliarm1}) as
\begin{eqnarray}
{\cal L}_{F}&=&- \left(
\vert F_{\phi_{L}} \vert^{2}+ \vert F_{\phi_{R}} \vert^{2}+ 
\vert F_{\phi^{\prime}_{L}} 
\vert^{2}+ 
\vert F_{\phi^{\prime}_{R}} 
\vert^{2}+
\vert F_{S} \vert^{2} 
\right) 
\nonumber \\
&=&
\vert \mu_{L} \vert^{2} 
\vert \phi^{\prime}_{L} \vert^{2}+
\vert \mu_{R} \vert^{2}
\vert \phi^{\prime}_{R} \vert^{2}+
\vert \mu_{L} \vert^{2} 
\vert \phi_{L} \vert^{2}+
\vert \mu_{R} \vert^{2} 
\vert \phi_{R} \vert^{2}+
\vert \mu_{S} \vert^{2} 
\vert S \vert^{2}.
\label{ftoscalarpotential}
\end{eqnarray}
If we perform the same program to $D$-fields we get
\begin{eqnarray}
D^{i}_{L}&=&- \frac{g_{L}}{2} \left[  
\bar{\phi}_{L}\sigma^{i}
\phi_{L}+
\bar{\phi}^{\prime}_{L}
\sigma^{i}\phi^{\prime}_{L}
\right] \,\ , \nonumber \\
D^{i}_{R}&=&- \frac{g_{R}}{2} \left[  
\bar{\phi}_{R}\sigma^{i}
\phi_{R}+
\bar{\phi}^{\prime}_{R}
\sigma^{i}\phi^{\prime}_{R}
\right] \,\ , \nonumber \\
D^{BL}&=&- \frac{g_{BL}}{2} \left[  
- \vert \phi_{L}\vert^{2}-
\vert \phi_{R}\vert^{2}+
\vert \phi^{\prime}_{L}
\vert^{2}+
\vert \phi^{\prime}_{R}
\vert^{2}+
\vert S \vert^{2}
\right] \,\ ,
\end{eqnarray}
According with Eq.(\ref{auxiliarm1})
\begin{eqnarray}
{\cal L}_{D}&=&- \frac{1}{2} \left( D^{i}_{L}D^{i}_{L}+
D^{i}_{R}D^{i}_{R}+
D^{BL}D^{BL} \right) 
\nonumber \\
&=& 
- \frac{g^{2}_{L}}{8} \left[
\left(
\vert \phi_{L}\vert^{2}-
\vert \phi^{\prime}_{L}\vert^{2}
\right)^{2}+4
\vert 
\phi^{\dagger \,\ \prime}_{L} 
\phi_{L}
\vert^{2}
\right] -
\frac{g^{2}_{R}}{8} \left[
\left(
\vert \phi_{R}\vert^{2}-
\vert \phi^{\prime}_{R}\vert^{2}
\right)^{2}+4
\vert 
\phi^{\dagger \,\ \prime}_{R} 
\phi_{R}
\vert^{2}
\right] 
\nonumber \\
&-& 
\frac{g^{2}_{BL}}{8}\left\{
\left[ \left( 
\vert \phi_{L}\vert^{2}-
\vert \phi^{\prime}_{L}\vert^{2}
\right)+
\left( 
\vert \phi_{R}\vert^{2}-
\vert \phi^{\prime}_{R}\vert^{2}
\right) \right]^{2}+ 
\left( \vert S \vert^{2}\right)^{2}-2
\left( 
\vert \phi_{L}\vert^{2}-
\vert \phi^{\prime}_{L}\vert^{2}
\right) \vert S \vert^{2}
\right. \nonumber \\
&-& \left. 2 
\left( 
\vert \phi_{R}\vert^{2}-
\vert \phi^{\prime}_{R}\vert^{2}
\right) \vert S \vert^{2}
\right\},
\label{dtoscalarpotential}
\end{eqnarray}
we have used the relation
\begin{equation}
\vec{\sigma}_{AB}\cdot 
\vec{\sigma}_{CD}=
2 \delta_{AD}\delta_{BC}-
\delta_{AB}\delta_{CD}.
\end{equation}

\section{Scalar in MSUSM.}

Calculating Eq.(\ref{ep1}) with the help of 
Eqs.(\ref{develop},\ref{constraintequationsinmodel1}) and 
using as base the following set of 
scalars 
$H_{\phi_{L}},H_{\phi_{R}},
H_{\phi^{\prime}_{L}},
H_{\phi^{\prime}_{R}},
H_{S}$, the mass matrix, with the help of 
Eq.(\ref
{constraintequationsinmodel1}), 
the mass matrix is written as
\begin{eqnarray}
{\cal M}^{2}_{CP-even} &=&\left( 
\begin{array}{ccccc}
{\bf {\cal M}^{even}_{11}} & 
{\bf {\cal M}^{even}_{12}} & 
{\bf {\cal M}^{even}_{13}} & 
{\bf {\cal M}^{even}_{14}} & 
{\bf {\cal M}^{even}_{15}} \\
{\bf {\cal M}^{even}_{12}} & 
{\bf {\cal M}^{even}_{22}} & 
{\bf {\cal M}^{even}_{23}} & 
{\bf {\cal M}^{even}_{24}} & 
{\bf {\cal M}^{even}_{25}} \\
{\bf {\cal M}^{even}_{13}} & 
{\bf {\cal M}^{even}_{23}} & 
{\bf {\cal M}^{even}_{33}} & 
{\bf {\cal M}^{even}_{34}} & 
{\bf {\cal M}^{even}_{35}} \\
{\bf {\cal M}^{even}_{14}} & 
{\bf {\cal M}^{even}_{24}} & 
{\bf {\cal M}^{even}_{34}} & 
{\bf {\cal M}^{even}_{44}} & 
{\bf {\cal M}^{even}_{45}} \\
{\bf {\cal M}^{even}_{15}} & 
{\bf {\cal M}^{even}_{25}} & 
{\bf {\cal M}^{even}_{35}} & 
{\bf {\cal M}^{even}_{45}} & 
{\bf {\cal M}^{even}_{55}}
\end{array}
\right)
\end{eqnarray}
where we have defined:
\begin{eqnarray}
{\bf {\cal M}^{even}_{11}}&=&
\left( 
\frac{g^{2}_{L}+g^{2}_{BL}}{4}
\right)v^{2}_{L}+ 
\beta^{2}_{\phi_{L}} 
\frac{v^{\prime}_{L}}{v_{L}}
, \,\
{\bf {\cal M}^{even}_{12}}= 
\frac{g^{2}_{BL}}{4}v_{L}v_{R}
, \nonumber \\
{\bf {\cal M}^{even}_{13}}&=&- 
 \beta^{2}_{\phi_{L}}+
\left( 
\frac{g^{2}_{L}+g^{2}_{BL}}{4}
\right) v_{L}v^{\prime}_{L}
, \,\
{\bf {\cal M}^{even}_{14}}=- 
\frac{g^{2}_{BL}}{4}v_{L}
v^{\prime}_{R}
, \,\
{\bf {\cal M}^{even}_{15}}
=-2xv_{L}
, \nonumber \\
{\bf {\cal M}^{even}_{22}}&=& 
\left( 
\frac{g^{2}_{R}+g^{2}_{BL}}{4}
\right)v^{2}_{R}+ 
\beta^{2}_{\phi_{R}} 
\frac{v^{\prime}_{R}}{v_{R}}
, \,\
{\bf {\cal M}^{even}_{23}}= 
\frac{g^{2}_{BL}}{4}v_{R}
v^{\prime}_{L}
, \nonumber \\
{\bf {\cal M}^{even}_{24}}&=&- 
\beta^{2}_{\phi_{R}}- 
\left( 
\frac{g^{2}_{R}+g^{2}_{BL}}{4}
\right)v_{R}v^{\prime}_{R}
, \,\
{\bf {\cal M}^{even}_{25}}
=-2xvv_{R}
, \nonumber \\
{\bf {\cal M}^{even}_{33}}&=& 
\left( 
\frac{g^{2}_{L}+g^{2}_{BL}}{4}
\right)v^{\prime 2}_{L}+ 
\beta^{2}_{\phi_{L}} 
\frac{v_{L}}{v^{\prime}_{L}}
, \,\
{\bf {\cal M}^{even}_{34}}= 
\frac{g^{2}_{BL}}{4}
v^{\prime}_{L}v^{\prime}_{R}
, \,\
{\bf {\cal M}^{even}_{35}}=2x
v^{\prime}_{L}
, \nonumber \\
{\bf {\cal M}^{even}_{44}}&=&
\left( 
\frac{g^{2}_{R}+g^{2}_{BL}}{4}
\right)v^{\prime 2}_{R}+ 
\beta^{2}_{\phi_{R}} 
\frac{v_{R}}{v^{\prime}_{R}}
, \,\
{\bf {\cal M}^{even}_{45}}=2x
v^{\prime}_{R}, \nonumber \\
{\bf {\cal M}^{even}_{45}}&=&2x^{2}.
\end{eqnarray}

It is easy to show the following results
\begin{eqnarray}
{\mbox Det}
{\cal M}^{2}_{CP-even}&=&
\frac{1}{8v_{L}v_{R}v^{\prime}_{L}
v^{\prime}_{R}}\left\{
\beta^{2}_{\phi_{L}}
\beta^{2}_{\phi_{R}} \left[
\left( g^{2}_{BL}-8
\right)g^{2}_{R}+
\left( g^{2}_{BL}-8+g^{2}_{R}
\right)g^{2}_{L}\right] \right. \nonumber \\
&\times& \left. 
\left( 
v^{2}_{L}+v^{\prime 2}_{L}
\right)
\left( 
v^{\prime 2}_{R}-v^{2}_{R}
\right)x^{2}
\right\}, \nonumber \\
{\mbox Tr}
{\cal M}^{2}_{CP-even}&=& 
\left( 
\frac{g^{2}_{L}+g^{2}_{BL}}{4}
\right) \left(
v^{2}_{L}+v^{\prime 2}_{L}
\right) +
\left( 
\frac{g^{2}_{R}+g^{2}_{BL}}{4}
\right) \left(
v^{2}_{R}+v^{\prime 2}_{R}
\right) \nonumber \\
&+& 
\beta^{2}_{\phi_{L}} \left( 
\frac{v_{L}}{v^{\prime}_{L}} 
+ 
\frac{v^{\prime}_{L}}{v_{L}} 
\right)
+ 
\beta^{2}_{\phi_{R}} \left( 
\frac{v_{R}}{v^{\prime}_{R}} 
+ 
\frac{v^{\prime}_{R}}{v_{R}} 
\right)
+2x^{2}
. \nonumber \\
\label{dettrcp-even}
\end{eqnarray}
This matrix has no Goldstone bosons and five mass eigenstates, which we 
denote as $H^{0}_{1},H^{0}_{2},
H^{0}_{3},H^{0}_{4},
H^{0}_{5}$.

We can using Eq.(\ref{dettrcp-even}), get 
the following expression for our light CP-even boson
\begin{eqnarray}
M_{H^{0}_{1}}&=&
\sqrt{\left( 
\frac{g^{2}_{L}+g^{2}_{BL}}{4}
\right) \left(
v^{2}_{L}+v^{\prime 2}_{L}
\right) +
\left( 
\frac{g^{2}_{R}+g^{2}_{BL}}{4}
\right) \left(
v^{2}_{R}+v^{\prime 2}_{R}
\right)} \nonumber \\
&=&
\sqrt{\left( 
\frac{g^{2}_{L}+g^{2}_{BL}}{4}
\right) v^{2}_{L} \left(
1+ \tan^{2} \beta
\right) +
\left( 
\frac{g^{2}_{R}+g^{2}_{BL}}{4}
\right) v^{2}_{R} \left(
1+ \tan^{2} \alpha
\right)},
\label{hlmassexpression}
\end{eqnarray}
where we have used Eq.(\ref{defbetaalpha}).

\section{Pseudoscalar in MSUSM.}

On this case using the base given by 
$F_{\phi_{L}},F_{\phi_{R}},
F_{\phi^{\prime}_{L}},
F_{\phi^{\prime}_{R}},
F_{S}$, the mass matrix, with the help of 
Eq.(\ref
{constraintequationsinmodel1}), 
the mass matrix is written as
\begin{eqnarray}
{\cal M}^{2}_{CP-odd} &=&\left( 
\begin{array}{ccccc}
{\bf {\cal M}^{odd}_{11}} & 0 & 
{\bf {\cal M}^{odd}_{13}} & 0 & 0 \\
0 & {\bf {\cal M}^{odd}_{22}} & 0 & 
{\bf {\cal M}^{odd}_{24}} & 0 \\
{\bf {\cal M}^{odd}_{13}} & 0 & 
{\bf {\cal M}^{odd}_{33}} & 0 & 
0 \\
0 & {\bf {\cal M}^{odd}_{24}} & 0 & 
{\bf {\cal M}^{odd}_{44}} & 0 \\
0 & 0 & 0 & 0 & 
{\bf {\cal M}^{odd}_{55}}
\end{array}
\right)
\end{eqnarray}
where we have defined:
\begin{eqnarray}
{\bf {\cal M}^{odd}_{11}}&=& 
\beta^{2}_{\phi_{L}} 
\frac{v^{\prime}_{L}}{v_{L}}, 
\,\ 
{\bf {\cal M}^{odd}_{13}}= 
\beta^{2}_{\phi_{L}}, \,\ 
{\bf {\cal M}^{odd}_{22}}= 
\beta^{2}_{\phi_{R}} 
\frac{v^{\prime}_{R}}{v_{R}}, 
\,\ 
{\bf {\cal M}^{odd}_{24}}= 
\beta^{2}_{\phi_{L}}, 
\nonumber \\
{\bf {\cal M}^{odd}_{33}}
&=&\beta^{2}_{\phi_{L}} 
\frac{v_{L}}{v^{\prime}_{L}}, 
\,\ 
{\bf {\cal M}^{odd}_{44}}
=\beta^{2}_{\phi_{R}} 
\frac{v_{R}}{v^{\prime}_{R}}, 
\,\ 
{\bf {\cal M}^{odd}_{55}}
=\beta^{2}_{\phi_{S}}.
\label{matpseum1}
\end{eqnarray}
Note that the basis $(F_{\phi_{L}},
F_{\phi^{\prime}_{L}})$ does not mix with $(F_{\phi_{R}},
F_{\phi^{\prime}_{R}}$ neither with $(F_{S})$.

It is easy to show the following results
\begin{eqnarray}
{\mbox Det}
{\cal M}^{2}_{CP-odd}&=&0, 
\nonumber \\
{\mbox Tr}
{\cal M}^{2}_{CP-odd}&=& 
4\beta^{2}_{\phi_{S}}+
\beta^{2}_{\phi_{L}}\left(
\tan \beta + \cot \beta
\right) +
\beta^{2}_{\phi_{R}}\left(
\tan \alpha + \cot \alpha
\right), \nonumber \\
\end{eqnarray}
where the characteristic equation is given by:
\begin{eqnarray}
x^{2}\left( 4\beta^{2}_{\phi_{S}}
\right) \left(
\beta^{2}_{\phi_{\phi_{L}}} 
(v^{2}_{L}+v^{\prime 2}_{L})- 
v_{L}v^{\prime}_{L}x \right) 
\left(
\beta^{2}_{\phi_{\phi_{R}}} 
(v^{2}_{R}+v^{\prime 2}_{R})- 
v_{R}v^{\prime}_{R}x \right)=0.
\end{eqnarray}
This mass matrix has two Goldstone bosons,$G^{0}_{1},G^{0}_{2}$ (they 
will become the longitudinal components of the $Z^{0}_{m}$ and 
$Z^{\prime 0}_{m}$ neutral vector bosons). 

We have also three mass 
eigenstates, 
$A^{0}_{1},A^{0}_{2},A^{0}_{3}$ 
and their masses are given by:
\begin{eqnarray}
m^{2}_{A^{0}_{1}}&=&
\beta^{2}_{\phi_{L}}\left(
\tan \beta + \cot \beta
\right) \gg M^{2}_{Z}, \nonumber \\
m^{2}_{A^{0}_{2}}&=&
\beta^{2}_{\phi_{R}}\left(
\tan \alpha + \cot \alpha
\right) \gg M^{2}_{Z^{\prime}}, \nonumber \\
m^{2}_{A^{0}_{3}}&=&4
\beta_{\phi_{S}}.
\label{eigenvaluescpodd}
\end{eqnarray}
see 
Eq.(\ref{neutralgaugebosonsmasses}). 
The pseudo-scalar scalar at MSSM has the following mass expression \cite{Rodriguez:2019mwf}:
\begin{eqnarray}
M^{2}_{A^{0}}&=& M^{2}_{12} \left( \tan \beta + \cot \beta \right),
\end{eqnarray}
and its values is similar expression for our $m^{2}_{A^{0}_{1}}$.

Therefore, the mass eigenstates can be defined as:
\begin{eqnarray}
\left( 
\begin{array}{c}
G^{0}_{1} \\
A^{0}_{1} \\
G^{0}_{2} \\
A^{0}_{2} \\
A^{0}_{3}
\end{array}
\right)
 &=&\left( 
\begin{array}{ccccc}
\cos \beta & \sin \beta & 
0 & 0 & 0 \\
- \sin \beta & \cos \beta 
& 0 & 0 & 0 \\
0 & 0 & 
\cos \alpha & \sin \alpha & 
0 \\
0 & 0 & 
- \sin \alpha & \cos \alpha 
& 0 \\
0 & 0 & 0 & 0 & 1
\end{array}
\right)
\left( 
\begin{array}{c}
F_{\phi_{L}} \\
F_{\phi^{\prime}_{L}} \\
F_{\phi_{R}} \\
F_{\phi^{\prime}_{R}} \\
F_{S}
\end{array}
\right).
\end{eqnarray}

\section{Charged fields in MSUSM.}

On this case the basis is given by 
$\phi^{\pm}_{L}, \phi^{\pm}_{R}, \phi^{\prime \pm}_{L}, \phi^{\prime \pm}_{R}$, with the help of 
Eq.(\ref{constraintequationsinmodel1}), we get
the mass matrix is written as
\begin{eqnarray}
{\cal M}^{2}_{Charged} &=&\left( 
\begin{array}{cccc}
{\bf {\cal M}^{charg}_{11}} & 0 & 
{\bf {\cal M}^{charg}_{13}} & 0 \\
0 & {\bf {\cal M}^{charg}_{22}} & 0 & 
{\bf {\cal M}^{charg}_{24}} \\
{\bf {\cal M}^{charg}_{13}} & 0 & 
{\bf {\cal M}^{charg}_{33}} & 0 \\
0 & {\bf {\cal M}^{charg}_{24}} & 0 & 
{\bf {\cal M}^{charg}_{44}} 
\end{array}
\right)
\end{eqnarray}
where we have defined:
\begin{eqnarray}
{\bf {\cal M}^{charg}_{11}}&=& 
\beta^{2}_{\phi_{L}} 
\frac{v^{\prime}_{L}}{v_{L}}+ \frac{g^{2}_{L}}{4}
v^{\prime 2}_{L}, 
\,\ 
{\bf {\cal M}^{charg}_{13}}= 
\beta^{2}_{\phi_{L}}+ \frac{g^{2}_{L}}{4}
v^{\prime}_{L}v_{L}, 
\nonumber \\ 
{\bf {\cal M}^{charg}_{22}}&=& 
\beta^{2}_{\phi_{R}} 
\frac{v^{\prime}_{R}}{v_{R}}+ \frac{g^{2}_{R}}{4}
v^{\prime 2}_{R}, 
\,\ 
{\bf {\cal M}^{charg}_{24}}= 
\beta^{2}_{\phi_{R}}+ \frac{g^{2}_{R}}{4}
v^{\prime}_{R}v_{R},  
\nonumber \\
{\bf {\cal M}^{charg}_{33}}
&=& 
\beta^{2}_{\phi_{L}} 
\frac{v_{L}}{v^{\prime}_{L}}+ \frac{g^{2}_{L}}{4}
v^{2}_{L}, 
\,\ 
{\bf {\cal M}^{charg}_{44}}
=\beta^{2}_{\phi_{R}} 
\frac{v_{R}}{v^{\prime}_{R}}+ \frac{g^{2}_{R}}{4}
v^{2}_{R}.
\label{matchar1m1} 
\end{eqnarray}
Note that the basis $(\phi^{\pm}_{L},
\phi^{\prime \pm}_{L})$ does not mix with $(\phi^{\pm}_{R},
\phi^{\prime \pm}_{R})$.

It is easy to show the following results
\begin{eqnarray}
{\mbox Det}
{\cal M}^{2}_{Charged}&=&0, 
\nonumber \\
{\mbox Tr}
{\cal M}^{2}_{Charged}&=& 
\left( \frac{\beta^{2}_{\phi_{L}}}
{v_{L}v^{\prime}_{L}}+ 
\frac{g^{2}_{L}}{4} \right)
\left(
v^{2}_{L}+ v^{\prime 2}_{L}
\right)+
\left( \frac{\beta^{2}_{\phi_{R}}}
{v_{R}v^{\prime}_{R}}+ 
\frac{g^{2}_{R}}{4} \right)
\left(
v^{2}_{R}+ v^{\prime 2}_{R}
\right), \nonumber \\
\end{eqnarray}
where the characteristic equation is given by:
\begin{eqnarray}
&x^{2}&\left[
4 \beta^{2}_{\phi_{\phi_{L}}} 
\left( v^{2}_{L}+v^{\prime 2}_{L} \right)+ 
v_{L}v^{\prime}_{L} \left(
g^{2}_{L} \left( v^{2}_{L}+v^{\prime 2}_{L} \right)
\right)-4x \right] 
\nonumber \\
&\times& 
\left[
4 \beta^{2}_{\phi_{\phi_{R}}} 
\left( v^{2}_{R}+v^{\prime 2}_{R} \right)+ 
v_{R}v^{\prime}_{R} \left(
g^{2}_{R} \left( v^{2}_{R}+v^{\prime 2}_{R} \right)
\right)-4x \right]=0.
\end{eqnarray}
This mass matrix has two Goldstone bosons,$G^{\pm}_{1},G^{\pm}_{2}$
(they will become the longitudinal components of the $W^{\pm}_{L}$ and $W^{\pm}_{R}$ charged vector bosons). 

We have also two mass 
eigenstates, $H^{\pm}_{1},H^{\pm}_{2}$ and their masses are given by:
\begin{eqnarray}
m^{2}_{H^{\pm}_{1}}&=&
\left( \frac{\beta^{2}_{\phi_{L}}}
{v_{L}v^{\prime}_{L}}+ 
\frac{g^{2}_{L}}{4} \right)
\left(
v^{2}_{L}+ v^{\prime 2}_{L}
\right) \gg M^{2}_{W_{L}}, \nonumber \\
m^{2}_{H^{\pm}_{2}}&=&
\left( \frac{\beta^{2}_{\phi_{R}}}
{v_{R}v^{\prime}_{R}}+ 
\frac{g^{2}_{R}}{4} \right)
\left(
v^{2}_{R}+ v^{\prime 2}_{R}
\right) \gg M^{2}_{W_{R}},.
\label{eigenvaluescharged}
\end{eqnarray}
see Eq.(\ref{chargedgaugebosonsmasses}).

Therefore, the mass eigenstates can be defined as:
\begin{eqnarray}
\left( 
\begin{array}{c}
G^{\pm}_{1} \\
H^{\pm}_{1} \\
G^{\pm}_{2} \\
H^{\pm}_{2} 
\end{array}
\right)
 &=&\left( 
\begin{array}{cccc}
\cos \beta & \sin \beta & 
0 & 0 \\
- \sin \beta & \cos \beta 
& 0 & 0 \\
0 & 0 & 
\cos \alpha & \sin \alpha \\
0 & 0 & 
- \sin \alpha & \cos \alpha 
\end{array}
\right)
\left( 
\begin{array}{c}
\phi^{\pm}_{L} \\
\phi^{\prime \pm}_{L} \\
\phi^{\pm}_{R} \\
\phi^{\prime \pm}_{R} 
\end{array}
\right).
\end{eqnarray}



\begin{thebibliography}{99}

\bibitem{Rodriguez:2019mwf} 
M. C. Rodriguez,
{\it The Minimal Supersymmetric Standard Model (MSSM) and General Singlet Extensions of the MSSM (GSEMSSM), a short review},
[arXiv:1911.13043 [hep-ph]].

\bibitem{Rodriguez:2016esw}M. C. Rodriguez and I. V. Vancea, 
{\it Flat Directions and Leptogenesis in a "New" $\mu \nu$SSM},
arXiv:1603.07979 [hep-ph].

\bibitem{Rodriguez:2020}M. C. Rodriguez, 
{\it Short review about the MSSM with three right-handed neutrinos (MSSM3RHN)}, arXiv:2003.04638 [hep-ph].

\bibitem{Diaz:2020qti}
H. Diaz, E. Castillo-Ruiz, O. P. Ravinez and V. Pleitez,
{\it Explicit parity violation in $SU(2)_L\otimes SU(2)_R\otimes U(1)_{B-L}$ models}, [arXiv:2002.03524 [hep-ph]].

\bibitem{Davidson:1987mh}
A. Davidson and K. C. Wali,
{\it Universal Seesaw Mechanism?},
{\sl Phys. Rev. Lett.}{\bf 59}, 393, (1987);
doi:10.1103/PhysRevLett.59.393.

\bibitem{Davidson:1987tr}
A. Davidson and K. C. Wali,
{\it Family Mass Hierarchy From Universal Seesaw Mechanism},
{\sl Phys. Rev. Lett.}{\bf 60}, 1813, (1988);
doi:10.1103/PhysRevLett.60.1813.

\bibitem{susylr} 
K. Huitu, J. Maalampi and M. Raidal, 
{\sl Nucl. Phys.}{\bf B420}, 449 (1994); 
C.S.Aulakh,A.Melfo and G.Senjanovic,
{\sl Phys.Rev.}{\bf D57},4174 (1998); 
G. Barenboim and N. Rius,
{\sl Phys. Rev.}{\bf D58}, 065010, (1998); 
N. Setzer and S. Spinner,
{\sl Phys. Rev.} {\bf D71}, 115010 (2005).

\bibitem{doublet} 
K. S. Babu.B. Dutta and R.N. Mohapatra, 
{\sl Phys.Rev.}{\bf D65}:016005, (2002).

\bibitem{Hati:2018tge} C. Hati, S. Patra, P. Pritimita and U. Sarkar,
{\it Neutrino Masses and Leptogenesis in Left-Right Symmetric Models: A Review From a Model Building Perspective},
{\sl Front. in Phys.}{\bf 6}, 19, (2018);
doi:10.3389/fphy.2018.00019.

\bibitem{banks}T. Banks, {\it Supersymmetry and the Quark Mass Matrix},
{\sl Nucl. Phys.}{\bf B303}, 172, (1988).
\bibitem{ma}E. Ma, 
{\it Radiative Quark and Lepton Masses Through Soft Supersymmetry Breaking},
{\sl Phys. Rev.}{\bf D39}, 1922, (1989).
\bibitem{cmmc} C. M. Maekawa and M. C. Rodriguez, 
{\it Masses of fermions in supersymmetric models}, 
{\sl JHEP}{\bf 04}, 031, (2006), [hep-ph/0602074].
\bibitem{cmmc1} C. M.Maekawa and M. C.Rodriguez,
{\it Radiative Mechanism to Light Fermion Masses in the MSSM}, 
{\sl JHEP}{\bf 0801}, 072, (2008), [arXiv:0710.4943 [hep-ph]].
\bibitem{global} M. C. Rodriguez, 
{\it Neutrino masses in a supersymmetric model with exotic right-handed neutrinos in global $\mathcal {Z_3}$ $\bigotimes$ (B-L) symmetry},
{\it Int. J. Mod. Phys.}{\bf A36}, 2150010 (2021), 
arXiv:2007.14154 [hep-ph].


\bibitem{Zyla:2020zbs}
P. A. Zyla \textit{et al.} [Particle Data Group],
{\it Review of Particle Physics},
{\sl Prog. Theor. Exp. Phys.} {\bf 2020}, 083C01, (2020).

\end{thebibliography}
\end{document}